\documentclass[fleqn,usenatbib]{mnras}

\usepackage{newtxtext,newtxmath}

\usepackage[T1]{fontenc}
\usepackage{ae,aecompl}

\usepackage{graphicx}	
\usepackage{amsmath}	
\usepackage[utf8]{inputenc}

\title{Investigation of the Connection Between X-ray Binaries and Compact Star Clusters in NGC 628}

\author[S. Avdan et al.]{S. Avdan,$^{1}$\thanks{E-mail:kayaci.s@gmail.com} A. Akyuz,$^{1,2}$ 
S. Acar,$^{3}$
I. Akkaya Oralhan,$^{3}$
S. Allak,$^{1}$ \and
N. Aksaker,$^{1,4}$ 
\\ \\
${^1}$Space Science and Solar Energy Research and Application Center (UZAYMER), University of Çukurova, 01330, Adana, Turkey\\
$^2$Department of Physics, University of Çukurova, 01330, Adana, Turkey\\
$^3$Department of Astronomy and Space Sciences, Erciyes University, 38039, Kayseri, Turkey\\
$^4$Adana Organised Industrial Zones Vocational School of Technical Science, University of Çukurova, 01410, Adana, Turkey\\
\\
}
\date{Submitted 2022.}

\usepackage{graphicx}

\begin{document}

\maketitle

\begin{abstract}
We investigated for a possible connection between the types of X-ray binaries (XRBs) and the properties of compact star clusters in the nearby galaxy NGC 628. Using {\it Chandra} archival data covering the years 2001-2018, 75 X-ray sources were detected within the field of view of observations. 
A total of 69 XRBs, one of which is an ultraluminous X-ray source (ULX), were found to be in the luminosity range of 3$\times10^{36}\leq L_{X}$ $\leq $ 2$\times10^{39}$ erg s$^{-1}$. We identified the optical counterpart(s) of 15 of the 42 XRBs that coincide with the {\it HST} field of view via improved astrometry. We classified 15 of them as HMXBs based on the presence of the optical counterparts. The remaining sources with no optical counterparts were classified as LMXBs.
We also search compact star clusters in this galaxy using the multi-band optical images drawn from {\it HST} archives. 864 compact star clusters were identified and their ages and masses were determined by applying the best-fit SSP (Simple Stellar Population) model to their color-color diagram. We found that in NGC 628, HMXBs are associated with younger star clusters and LMXBs with older ones. Our findings support a connection between different types of XRBs and cluster ages, already known to exist for other galaxies.

\end{abstract}


\section{Introduction}

An X-ray binary (XRB) consists of a compact object (a black hole or neutron star) and a donor star with an X-ray luminosity of L$_{X}$ $\sim$ 10$^{36-38}$ erg s$^{-1}$. XRBs are generally classified into two broad classes as high-mass (HMXB) and low-mass (LMXB) depending on the mass of their donor stars, M$_{d}$. HMXBs have donor stars with M$_{d} \geq 10$ M$\odot$, while LMXBs are generally considered to have M$_{d} \leq$ 1 M$\odot$ \citep{2006csxs.book..623T}. However, as reported in the recent studies by \cite{2019ApJ...871..122J, 2020ApJ...890..150C, 2021ApJ...912...31H}, the classification can be further extended to the following M$_{d}$ ranges: M$_{d}$> 8 M$\odot$ for an HMXB, 3 M$\odot$< M$_{d}$ <8 M$\odot$ for an intermediate XRB (IMXB) and M$_{d}$<3 M$\odot$ for an LMXB. 

The X-ray emission from accreting XRBs exhibits strong, periodic or quasi-periodic variability on various time-scales. Their spectral states and accretion rates typically change from weeks to days, while light curves vary widely on all timescales above milliseconds \citep{2006ARA&A..44...49R}. 
Using data from both {\it Chandra} and {\it XMM-Newton} allow us to further probe XRBs in external galaxies. Reported results on XRBs in different galaxies highly resemble the spectral and temporal-spectral behavior of Galactic XRBs, including soft and hard spectral states \citep{2006ARA&A..44..323F, 2006MNRAS.371.1903I,2013A&A...555A..65H,2018ApJ...864..150V,2018A&A...620A..28S}.

Star clusters with high stellar density are favorite regions for the formation of XRBs. Since stellar encounters in such dense clusters can be high, this may increase the production of XRBs. While some XRBs that are spatially consistent with star clusters support this possibility, there are some that are spatially not matched. The latter suggests that, after their formation, they may have been ejected with sufficient velocities from their parent clusters \citep{2002ApJ...577..710Z, 2004MNRAS.348L..28K, 2011ApJ...741...86R, 2012ApJ...755...49G}.

The compact nature of individual star clusters in distant galaxies (beyond the local group) makes them difficult to find or distinguish; however, the high imaging capability of Hubble Space Telescope ({\it HST}) allows a detailed study of the extragalactic star clusters in galaxies with different morphological types \citep{2007AJ....133.1067W, 2009ApJ...704..453F, 2009A&A...501..949M, 2010ApJ...711.1263C, 2013ApJ...766...20L, 2016ApJ...824...71C}. 

Determining the physical characteristics of parent star clusters such as age, mass and metal abundance is important for understanding nature of XRBs. The roles of cluster ages, locations and also the constraints on the mass of donor stars for the classification of XRBs as HMXB or LMXBs, have been discussed in the literature for galaxies at various distances.  \citep{2004MNRAS.348L..28K, 2011ApJ...741...86R, 2019ApJ...871..122J, 2020ApJ...890..150C}.

In the present work, we focus on a possible connection between XRBs and compact star clusters in NGC 628, a late-type nearby spiral galaxy at a distance of 9.7 Mpc \citep{1988Sci...242..310T}. A number of studies have been reported on the search for star clusters in NGC 628 and on the determination of the luminosity function of the clusters identified in the {\it HST} images \citep[and references therein]{2002AJ....124.1393L, 2004A&A...416..537L, 2014AJ....147...78W, 2015ApJ...815...93G, 2017ApJ...841..131A,2017ApJ...841...92R}. In last two works (both in 2017), a comprehensive young star cluster catalog was produced using homogeneous imaging data (from UV to NIR) provided by the Legacy ExtraGalactic UV Survey (LEGUS), designed as a Hubble treasure program. The physical properties of these clusters were also presented in detail. We note that this catalog only includes star clusters of the galaxy designated as the eastern and central fields. However, the footprint of NGC 628 cover also the western region in the study of \cite{2022MNRAS.509..180L} where the luminosity function of their detected globular clusters (with ages >10 Gyr) was also derived by using {\it HST} observations.

In the present work, our goals are two fold: First, we will investigate the spatial connection between XRBs and compact star clusters for all three fields of this galaxy and the second, we will present a classification of these binaries by deploying archival data from {\it Chandra} and {\it HST}. For the classification, we considered the possible donor stars in XRBs and the ages of the associated compact clusters that possibly the host clusters.

The paper is organized as follows: X-ray/optical observations and data reductions are given in Section \ref{sec:obs}. X-ray source classification and properties of donor stars are presented in Section \ref{sec:x-ray} and identification of optical star clusters are in Section \ref{sec:optical}. In Section \ref{sec:discussion}, we summarize and discuss the results. Finally, in Section \ref{sec:summary}, are noted summary and conclusions from our main findings.

\section{Observations, Data Reductions and Analyses}
\label{sec:obs}

\subsection{{\it Chandra} Observations} \label{Sec.2.1}

NGC 628 was observed 13 times with {\it Chandra} between 2001$-$2018. The observations used in this study are detailed in Table \ref{T:x-ray}. The data reductions were performed using the {\scshape ciao} (Chandra Interactive Analysis of Observations) v4.12 software with the {\it Chandra} Calibration Database ({\scshape caldb}, v4.9.2.1). 

\begin{figure*}
\begin{center}
\includegraphics[angle=0,scale=0.25]{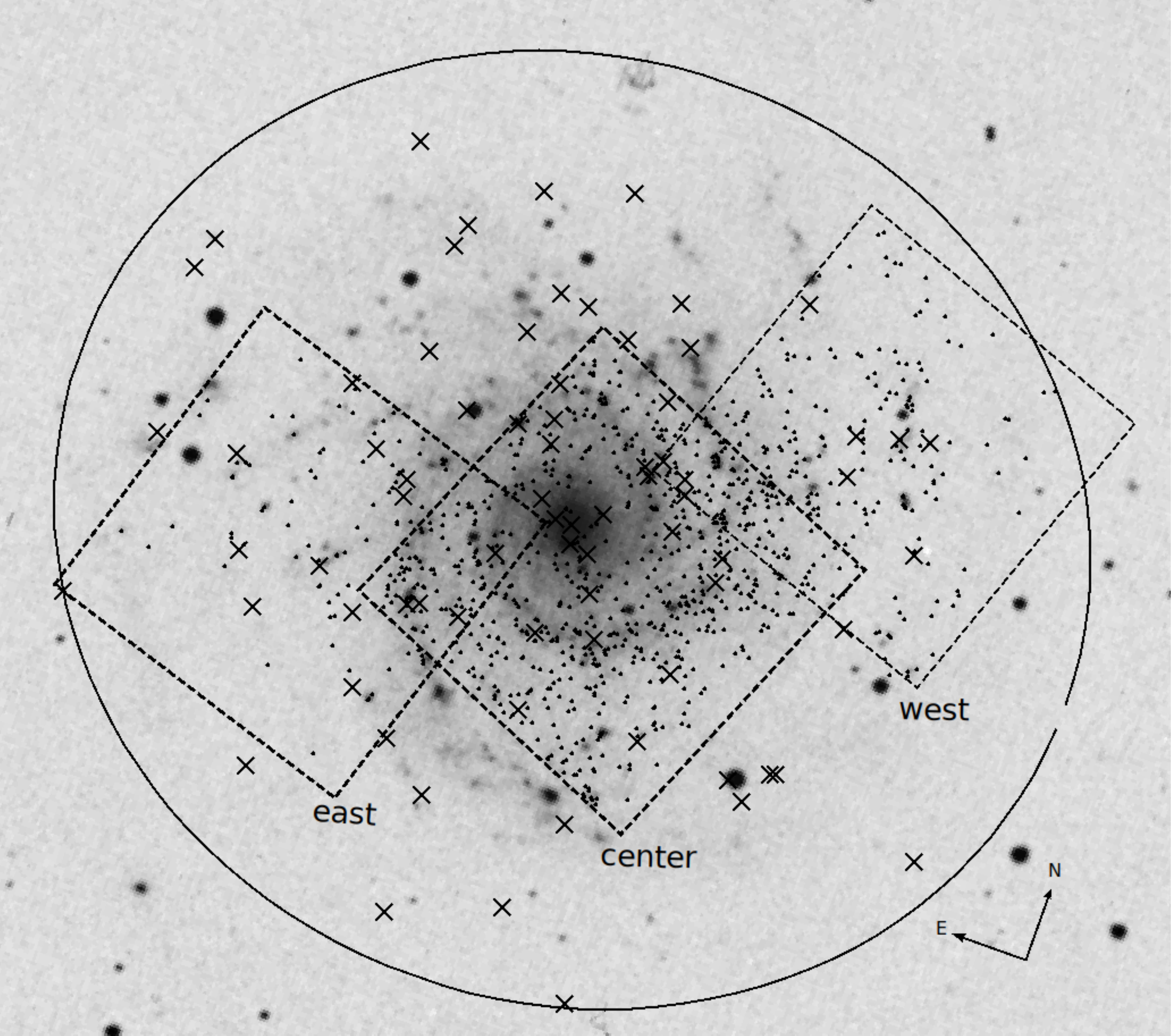}
\caption{DSS image of NGC 628. {\it HST}/ACS observation regions (east, center and west) which used in this study are shown as black dashed squares. The locations of detected star clusters are shown by black dots. Black crosses represent the position of X-ray sources within the D$_{25}$ region (black ellipse). } 
\label{F:dss}
\end{center}
\end{figure*}

In all observations used, the X-ray sources are located on the back-illuminated ACIS-S3 and the front-illuminated ACIS-S2 chips. The event files were created in four energy bands; soft (S) (0.3-1 keV), medium (M) (1-2 keV), hard (H) (2-8 keV) and the total (T) (0.3-8 keV). The source detection process was carried out using the {\scshape wavedetect} task in the total energy range with wavelet scales 2 and 4 pixels. A total of 75 X-ray sources were detected within the D$_{25}$ region of size 10\farcm5x9\farcm5 \citep{2017A&A...600A...6M}. Using available observations and the positions of these sources in {\it DSS} image are shown in Fig. \ref{F:dss}. We eliminated the sources outside this region since they are likely to be background objects. In addition, 6 more sources were removed from our list of sources in D$_{25}$ region; one is associated with the center of the galaxy, 2 sources are cataloged as AGNs \citep{2016yCat.9048....0M, 2021yCat.7290....0F}, and 3 sources are cataloged as SNRs as detailed by \cite{2010A&A...517A..91S}. We noticed that only 42 of these detected sources fall within the {\it HST} field of view. To estimate the count rates for each (detected) sources in 0.3-8 keV energy band, {\it specextract} task was used and for this, the source and background regions were mostly selected as 8$\arcsec$ radius circular regions. In case of faint and/or very close sources, these regions were reduced to circles 4$\arcsec$ $-$ 5$\arcsec$ in radius. The background regions were selected from the source-free regions on the same chip where the X-ray sources were located. If a detected source was not seen in another observation, we calculated an upper limit for the flux of the source at  90 \%  confidence level.

Spectral models were applied to the X-ray sources with >50 counts. Only 6 out of 42 sources were found to meet this criterion. All spectra were grouped into energy bins with minimum 10 counts and then spectral analyses were performed using the {\it XSPEC} v.12.11.0 package for the 0.3-8 keV energy range. One component models such as the power-law (po), a multi-color discblackbody (diskbb), black-body (bbody), and the two-component models such as po+diskbb and po+bbody were fitted the spectra with an absorption model (tbabs). We kept the absorption model parameter fixed to the Galactic column density of N$_{H}$ = 4.8$\times$ 10$^{20}$ cm$^{-2}$ \citep{1990ARA&A..28..215D}. The best-fit model parameters, chi-square ($\chi^{2}$) statistics and derived L$_{X}$ values with the adopted distance of 9.7 Mpc are given in Table \ref{T:x-ray_par}. Flux values for sources whose spectra cannot be fitted with the spectral models were calculated from the {\it webPIMMS} tool \footnote{https://cxc.harvard.edu/toolkit/pimms.jsp} using respective count rates in 0.3-8 keV. For this, we used the power-law model parameter of $\Gamma$=1.7 and the Galactic column density, N$_{H}$, for each source.


Examining the effect of the variability of XRBs will lead us to a more comprehensive understanding of their physics and classification. Therefore, we  investigated the long-term variability of the X-ray point sources in NGC 628 during the observed period of seventeen years. We calculated the ratio of the maximum (F$_{max}$) and minimum (F$_{min}$) flux in 0.3-8 keV energy range. This ratio is called as the variability factor and given by V$_{f}$= F$_{max}$/ F$_{min}$. We classified the X-ray point sources as variable sources when V$_{f}$>5. If a source shows high variability with V$_{f}$ > 10, we consider it to be a transient source. However, if 5 < V$_{f}$< 10 or V$_{f}$ < 5, the source was considered to be a transient candidate or unclear variable, respectively.


\subsection{{\it HST} Observations}

NGC 628 was observed in three different fields using the {\it HST}/ACS Wide Field Camera (WFC). {\it HST}/ACS observations cover center, east and west fields of the galaxy. These fields on the {\it DSS} image of the galaxy are shown in Fig. \ref{F:dss}. The log of observations are given in Table \ref{T:obs}. 

Photometric analyses of the archival {\it HST} data were performed using the {\it daophot} package in {\scshape iraf} v 2.16. Source detection was carried out using the {\it daofind} task in {\scshape iraf} by taking 3$\sigma$ detection threshold and FWHM (Full Width Half Maximum) of features 2.5 pixels. Using these parameters, $\sim$ 267000 optical sources were detected for the center, east and west fields in the F555W filter.

We performed aperture photometry for 0.5  and 3 pixels radii separately. For both of the 0.5 and 3 pixels radii, we took the background annulus to have a radius of 7 and a width of 1 pixel. We also obtained the concentration index (CI) values by using the difference in magnitude calculated within 0.5 and 3 pixels aperture radii, We used the zero point values from the \cite{2005PASP..117.1049S} to convert magnitudes to the Vega magnitudes (Vegamag).

Aperture photometry measures the flux within the user-defined aperture radius (3 pixels in our analysis). However,  the total flux is distributed over a wider area outside the aperture. In this case, an aperture correction is needed for user-defined aperture magnitude. This correction is defined as the difference of magnitudes between the measured in the user-defined aperture and the obtained in the standard photometric aperture (currently 5$\farcs$5).
We applied two different aperture correction methods. The first method, we performed the calculations with a radius between 0\farcs15 to 0\farcs5 for each filter using 20 isolated point sources which means there is no other bright source(s) within or very close to the selected aperture. As a second method, we were inspired an approach given by \cite{2000Aj..120..2938}.
We did not consider all detected sources as point sources due to their different shapes for the aperture correction. For this calculation the CI range was taken into account. For this, we first divided the sources into certain intervals according to their number density (see Table \ref{T:apcor}). Then, we selected at least 20 bright and isolated sources for each group. We obtained the aperture correction value for each filter given in Table \ref{T:apcor}.

After obtaining the aperture correction values of the sources, color-color diagrams were created for each method.  In the color-color diagram created according to the first method, we noticed that the sources are distributed over a wider area (called scattering) than the one created according to the second method. This indicates that there are a number of cluster candidates whose age cannot be determined by the related models (see sec. \ref{sec:ssp}). In other words, the first method could not represent all the sources in the diagram. Therefore, we used the second method for the aperture correction.


\subsection{Astrometry}

 

Identification of optical counterparts of the X-ray sources requires precise source positions. We compared the {\it Chandra} ACIS  images given in Table \ref{T:x-ray} and {\it HST}/ACS images given in Table \ref{T:obs} with the GAIA source catalogue for astrometric corrections of the sources. The {\scshape daofind} tool in {\scshape iraf} and {\it wavedetect} tool in {\scshape ciao} were used for the source detection in both {\it HST} and {\it Chandra} images, respectively. Initially, two well-matched X-ray sources (see Table \ref{T:astro}) were identified between Chandra and GAIA and astrometric calculations were performed following the study  of \cite{2022MNRAS.517.3495A}.
Then, using HST images for the three fields, reference sources matching GAIA and HST were determined. The coordinates and calculated offsets of these sources are given in Table 
\ref{T:astro}. Total astrometric errors between each comparison were found and also given in Table \ref{T:astro}. 
Finally, the positional error radius for center, east and west were derived as 0$\arcsec$.22, 0$\arcsec$.22 and 0$\arcsec$.27, respectively at 95\% confidence level combining all errors in quadrature.

Comparing X-ray source positions with {\it HST} images, 26 out of 42 XRBs are located in the central field, the remaining 10 of them are in the east field and 6 are in the west field in NGC 628. Taking into account the astrometric errors, we identify optical counterpart(s) for 15 XRBs. 10 of these are located in the central, 4 in the east and 1 in the west fields. Majority of them (13) have a unique optical candidate.

\section{X-ray sources}
\label{sec:x-ray}

\subsection{Classifications of the X-ray Sources}
 
Studies on source classification in the Milky Way and nearby galaxies show that X-ray sources are mostly of types HMXB and LMXB \citep{1989ARA&A..27...87F, 2003ApJ...595..719P, 2006csxs.book..623T, 2021ApJ...922..178R}. Determination of X-ray flux variabilities and spectral properties of XRBs help us to understand their nature. However, the low quality and insufficient statistics of available data (such as short exposures, faintness of sources) make it difficult to obtain the spectra of some sources. In such cases, an X-ray color-color diagram obtained for each source from their hardness ratios (HRs) will help us to distinguish the class of each source as an LMXB, a HMXB, a supernova remnant (SNR) or an absorbed source (mixture of types) \citep{2003ApJ...595..719P, 2019ApJ...879..112J}. For such an HR diagram, X-ray soft and hard color are defined as HR1=(M$-$S)/T and HR2=(H$-$M)/T, respectively. Here S, M, H and T are counts in the energy bands defined in Section \ref{Sec.2.1}.

The flux values for the 42 X-ray sources in the energy band 0.3-8 keV were determined to be in the range of 3$\times$10$^{-16}$ $-$ 2$\times$10$^{-13}$ erg cm$^{-2}$ s$^{-1}$ with corresponding X-ray luminosities in the range L$_{X}$ $\sim$ 3$\times$10$^{36}$ $-$ 2$\times$10$^{39}$ erg s$^{-1}$ for a distance of 9.7 Mpc. The derived lower limits of luminosity range is compatible with those of known bright XRBs \citep{2013ApJ...763...96V}. However our upper value is higher because one of the sources is an ultraluminous X-ray source (ULX) \citep{2005ApJ...630..228K}.
Therefore, all sources in our list can now be classified as XRBs. A color-color diagram obtained for the classification of these sources is given in Fig. \ref{F:2color}. Since some sources appear in several observations, we only used the observation that has their highest count rate to calculate the hard and soft colors. Using the method of 
\cite{2003ApJ...595..719P}, we classified 44$\%$ of the total source as LMXB, 30$\%$ as HMXB and 26$\%$ as others (soft or absorbed source).
 
As is known, classification of XRBs as LXMB and HMXB based on HR values is not certain enough to defined their true nature. For further clarification, we identified the optical counterpart(s) of these binaries using the HST/ACS optical images after the astrometric correction. Assuming that an optical counterpart in the positional uncertainty of an XRB is observed, it is likely the donor star. The mass of a donor star in an accreting binary systems is very useful in categorizing them as a LMXB or a HMXB. For this, we need to create Color-Magnitude Diagrams (CMDs). However, we are keeping in mind that optical emission of an XRB is thought to arise from either the accretion disc by reprocessing of X-rays in the outer disc region or from the  massive companion star or both \cite{2006MNRAS.371.1334R}.

\subsection{Color–Magnitude Diagram of Donor Stars}

\begin{figure}
\begin{center}
\includegraphics[angle=0,scale=0.40]{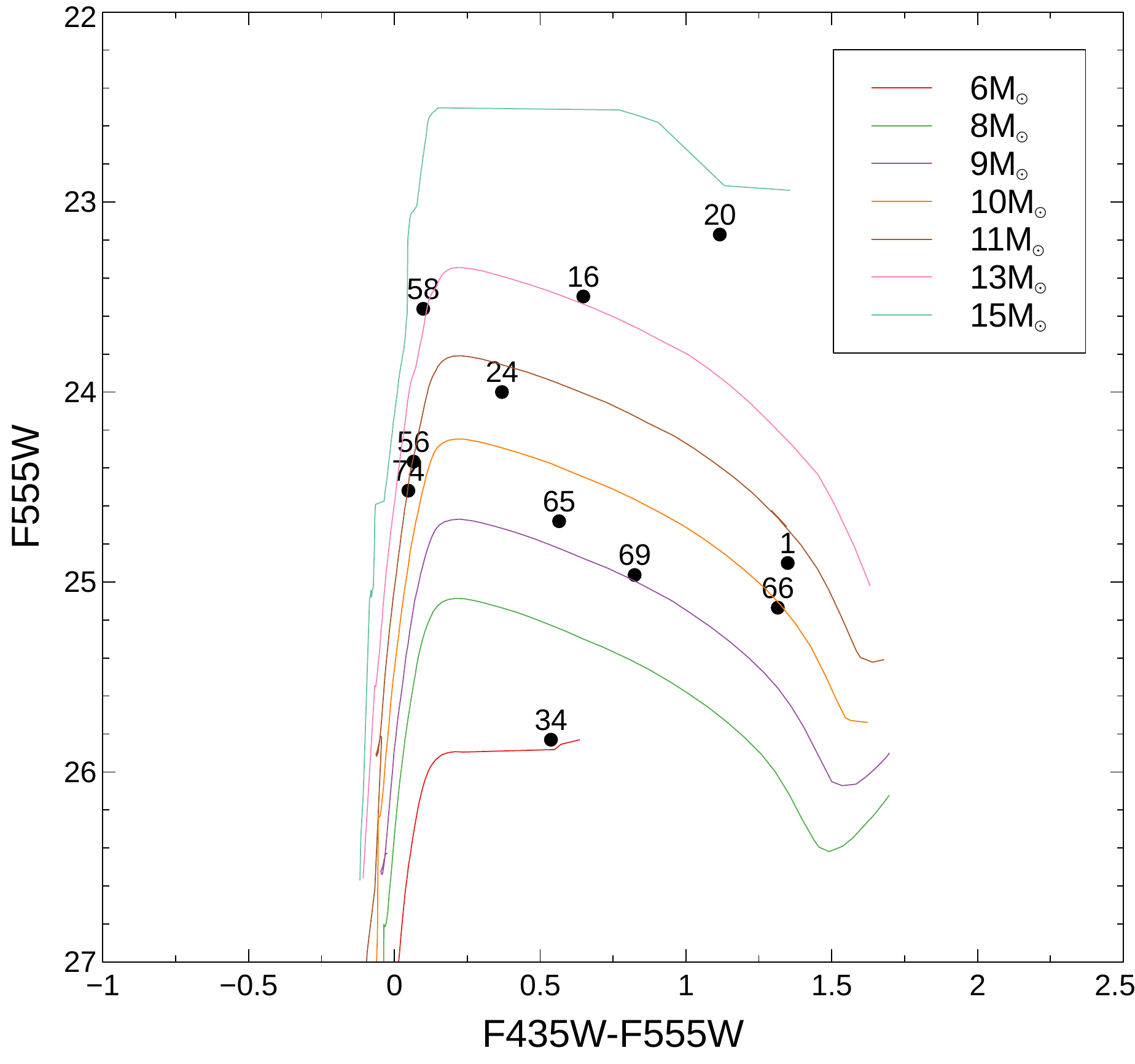}
\caption{The {\it HST} color-magnitude diagram (CMD) for potential donor stars (black dots ) which are unique counterparts of XRBs in NGC 628. These are compared with the Padova isochrones for different star masses. The theoretical isochrones have been corrected for extinction A$_{V}$= 0.46 mag. The numbers correspond to ID numbers of XRBs.}
\label{F:cmd}
\end{center}
\end{figure}

Optical identification of the donor star allows us to estimate its age, mass, and spectral type. Assuming that the observed optical emission originates from the donor star, the mass or age of the donor can be estimated from comparison of the generated CMDs with theoretical isochrones therefore, we obtained two CMDs (F435W-F555W versus F555W and F555W-F814W versus F555W) for XRBs that only have a single optical counterpart. We found very similar mass values for the most of the donors from these two CMDs so only one CMD is shown in Fig.\ref{F:cmd}. However, although the X5 and X22 have only one optical counterpart, they are not included in CMDs due to the optical candidates of the X5 and  X22 are very faint in the B and V bands, respectively. Parsec isochrones\footnote{Padova and Trieste Stellar Evolution Code; http://stev.oapd.inaf.it/cgi-bin/cmd} provided for the {\it HST}/ACS/WFC photometric systems with solar metallicity value were selected to construct the CMDs. The distance modulus was calculated as 29.93 mag and the isochrones have been corrected for reddening of E(B-V)=0.15 mag.


Considering that no consistent donor mass range exists in the literature, we adopt M$_{d}$ >8 M$_{\odot}$ for a HMXB; 3 M$_{\odot}$< M$_{d}$ <8 M$_{\odot}$ for an IMXB, and M$_{d}$<3 M$_{\odot}$ for an LMXB, in line with the recent studies by \cite{2019ApJ...871..122J, 2020ApJ...890..150C}. 
As seen in the Fig. \ref{F:cmd}, we found that the mass values of all candidate donor stars are in the range of 6M$_{\odot}$<M$_{d}$<15M$_{\odot}$. This indicates that these donor stars could possibly be classified as IMXB or HMXBs. Furthermore, we were not able to identify any optical counterparts for 27 XRBs sources, probably because they are too faint; see studies by \cite{2021ApJ...912...31H, 2021ApJ...922..178R} and references therein, who suggest an {\it HST} observation threshold for stars with a mass <3 M$_{\odot}$ at a distance of $\sim$ 10 Mpc.

\subsection{X-ray Luminosity Functions of XRBs}

X-ray luminosity functions (XLFs) are also be used to distinguish the types of XRBs. The studies showed that the XLFs of HMXBs scales with the star formation rate (SFR) of the host galaxy and the XLFs of LMXBs scales with stellar masses \citep{2003MNRAS.339..793G, 2004MNRAS.347L..57G, 2004MNRAS.349..146G, 2012MNRAS.419.2095M, 2014MNRAS.437.1698M, 2019ApJS..243....3L}. 

\begin{figure}
\begin{center}
\includegraphics[angle=0,scale=0.46]{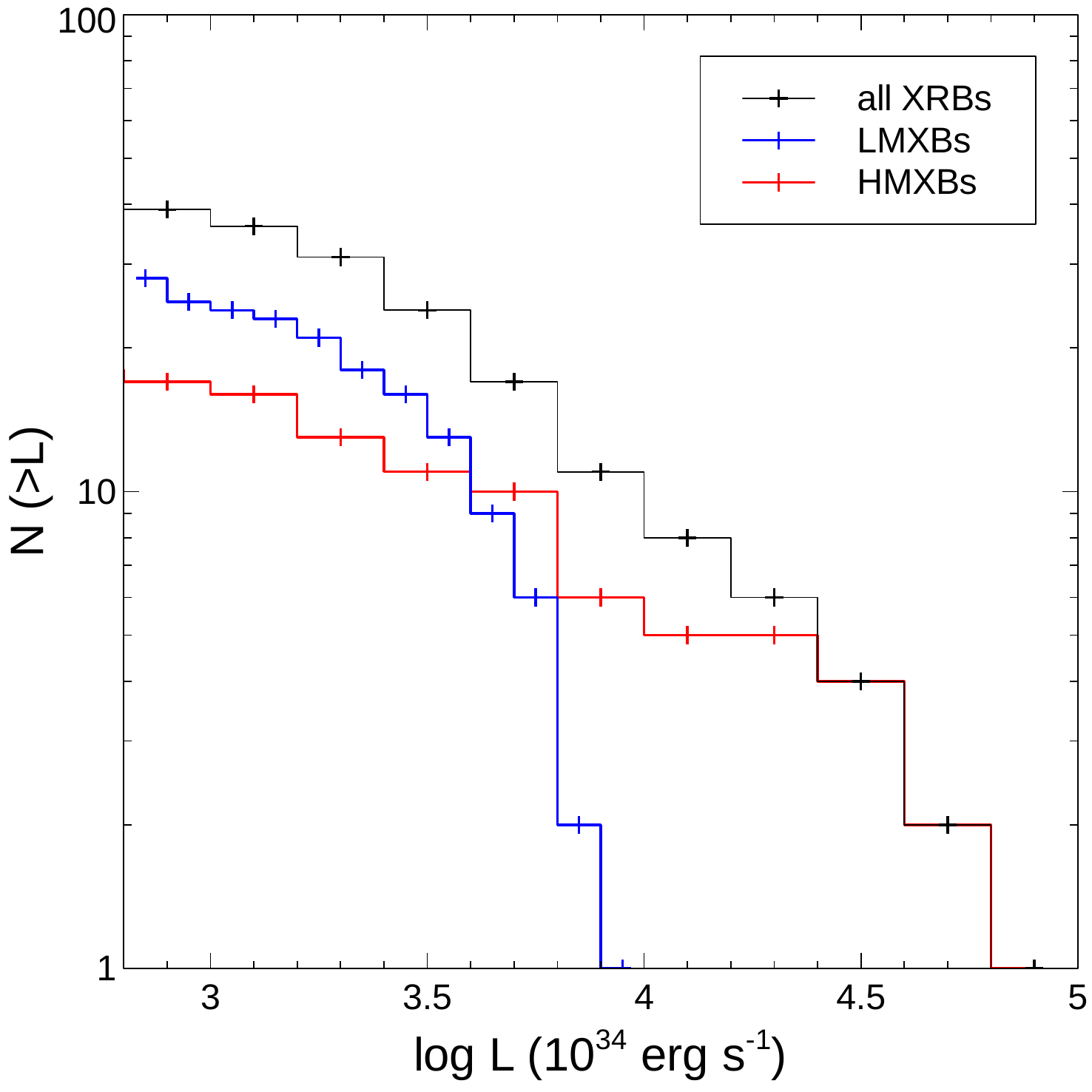}
\caption{The cumulative X-ray luminosity functions of XRBs in NGC 628. The blue, red and black lines represent LMXBs, HMXBs and all XRBs (LMXB+HMXB).} 
\label{F:cumulative}
\end{center}
\end{figure}

We obtained the X-ray luminosity functions for LMXBs and HMXBs in NGC 628 which were classified by searching optical counterparts. Since we have classified sources via optical counterparts, there is no need to make assumptions about scaling HMXBs with SFRs or LMXBs with stellar masses \citep{2020ApJ...890..150C, 2021ApJ...912...31H}.
The XLFs for LMXB, HMXB and both types are given in Fig. \ref{F:cumulative}.
\cite{2020ApJ...890..150C} examined the XLFs of XRBs in M101 galaxy which were classified as HMXBs and LMXBs via optical counterparts and they found that their XLFs exhibit different power-law slopes. Generally XLFs are best fitted with a power-law model and the XLFs of HMXBs extend to higher luminosities and flatter than the XLF of LMXBs \citep{2003ChJAS...3..257G, 2012MNRAS.419.2095M, 2020ApJ...890..150C}. 

As seen in Fig. \ref{F:cumulative}, there appears to be a difference in the shapes of the XLFs of HMXBs and LMXBs. Both distributions fit well with a simple power-law model but the difference in the slopes is notable, $\alpha$=-1.50$\pm0.07$ for HMXBs and $\alpha$=-1.70$\pm0.15$ for LMXBs. As expected, HMXBs range up to higher luminosity and have flatter slopes than the LMXBs. Also, XLFs of all XRBs in our sample well described by power-law model with $\alpha$ = -1.94$\pm0.10$. These results are consistent with \cite{2019ApJS..243....3L} despite using different source identification methods.

\section{Optical Star Clusters}
\label{sec:optical}

\subsection{Cluster Selection}

\begin{figure}
\begin{center}
\includegraphics[angle=0,scale=0.46,angle=0]{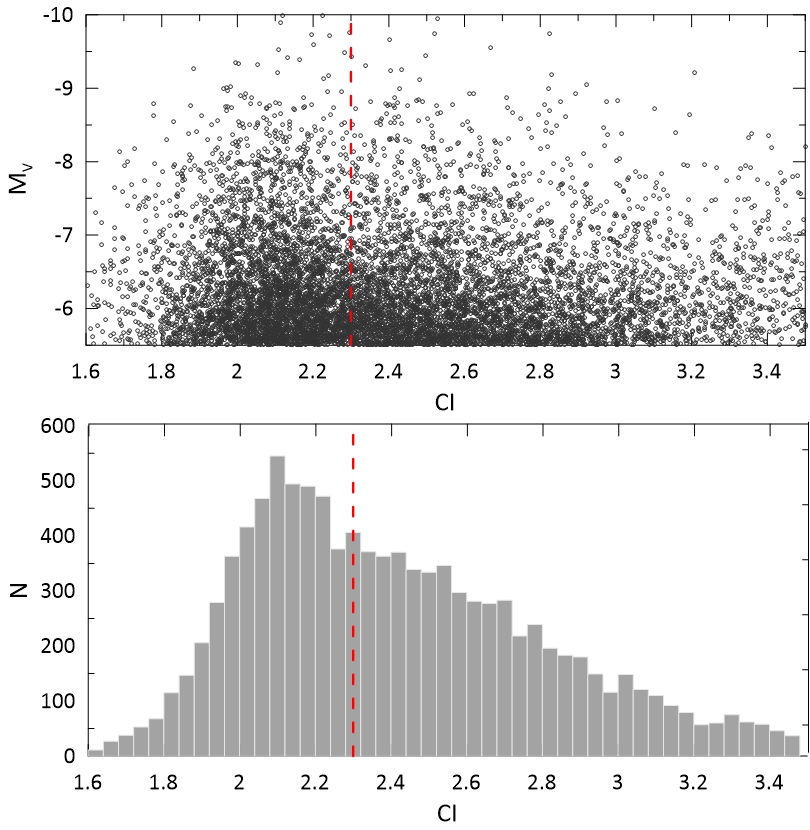}
\caption{(Upper) Distribution of sources brighter than M$_{V}\leq$ -5.5 according to CI. Vertical red line represents CI=2.3 to distinguish compact objects and stars. (Lower) Histogram of the CI values. Vertical red line represents CI=2.3 } 
\label{F:CI-fwhm}
\end{center}
\end{figure}

Star clusters are observed as compact objects in distant galaxies since the individual stars within these clusters can not be distinguished. Therefore several photometric criteria are imposed to define compact star clusters in these galaxies \citep{2010ApJ...719..966C, 2012MNRAS.419.2606B,2012ApJ...752...95J}. One of the criteria is the magnitude limit used to distinguish between stars and clusters due to the clusters are usually brighter than stars. As a first step, we choose magnitude of the sources as m$_{F555W}$ <25 mag. The second step is to apply CI to further distinguish clusters from stars. The CI values obtained for the sources versus the absolute magnitude in the V band (M$_{V}$) is shown in the upper panel of Fig. \ref{F:CI-fwhm} and the histogram of CI is given in the lower panel.
We interpret that sources have CI<2.3 to be defined as stars. On the other hand, CI>2.3 could be compact cluster candidates. Therefore, the sources that have CI>2.3 were taken for further steps. The third step is to calculate the size of the sources using the BAOlab/{\it Ishape} \citep{1999A&A...345...59L} software. {\it Ishape} convolves analytic profiles with the point-spread function (PSF), and calculates the best fit profile for each source. PSF was generated from bright isolated stars in the F555W filter for each field. We applied several profiles for the remaining sources and found the best fit profile as EFF (Elson-Fall-Freeman, \cite{1987ApJ...323...54E}) with a power index of 1.5. We also used 4 pixels as an input parameter (FITRAD) during the fitting procedure in {\it Ishape} task. The one of outputs is the FWHM that helps to find the size of the sources. If the FWHM > 0.2 pixels, they are defined as our cluster candidates.  We also converted the FWHM values to the effective radii (radius containing one-half of the total source light, r$_{\mathrm{eff}}$) by multiplying 1.13 as given in {\it ishape} manual \citep{1999A&AS..139..393L}. In the study of \cite{2004A&A...416..537L} on star clusters in NGC 628, the mean value of r$_{\mathrm{eff}}$ for 30 clusters is 3.65$\pm$0.55 pc from {\it ishape}. \cite{2017ApJ...841...92R} identified a total of 320 star clusters and found the sizes of them as 2.9 pc from {\it galfit}. In our study, this value is  $\sim$ 3.8 pc from {\it ishape}.


The final step is to select the isolated cluster candidates that do not have any object in the nearby  within a radius of 5 pixels. In addition, a final eye inspection was also made. As a result, a total of 970 clusters were identified in the east, center and west fields. The properties of a list of sample identified star clusters are given in Table \ref{T:agemass}. 

In general, when we compare our cluster identification criteria with studies in the literature involving cluster identification in NGC 628 such as \cite{2015ApJ...815...93G, 2017ApJ...841..131A,2017ApJ...841...92R}, we realized that several different criteria affect the number of clusters detected.These differences are as follows; the west field is not included in their works, the CI value was considered as 1.4 for center and 1.3 for west field and M$_{V}$ values are slightly different. In addition, only young (<200 Myr) and massive (5×10$^{3}$ M$\odot$) star clusters were selected in the study of \cite{2017ApJ...841...92R}.

\subsection{Determination of Age and Mass of Clusters}
\label{sec:ssp}

\begin{figure}
\begin{center}
\includegraphics[angle=0,scale=0.38]{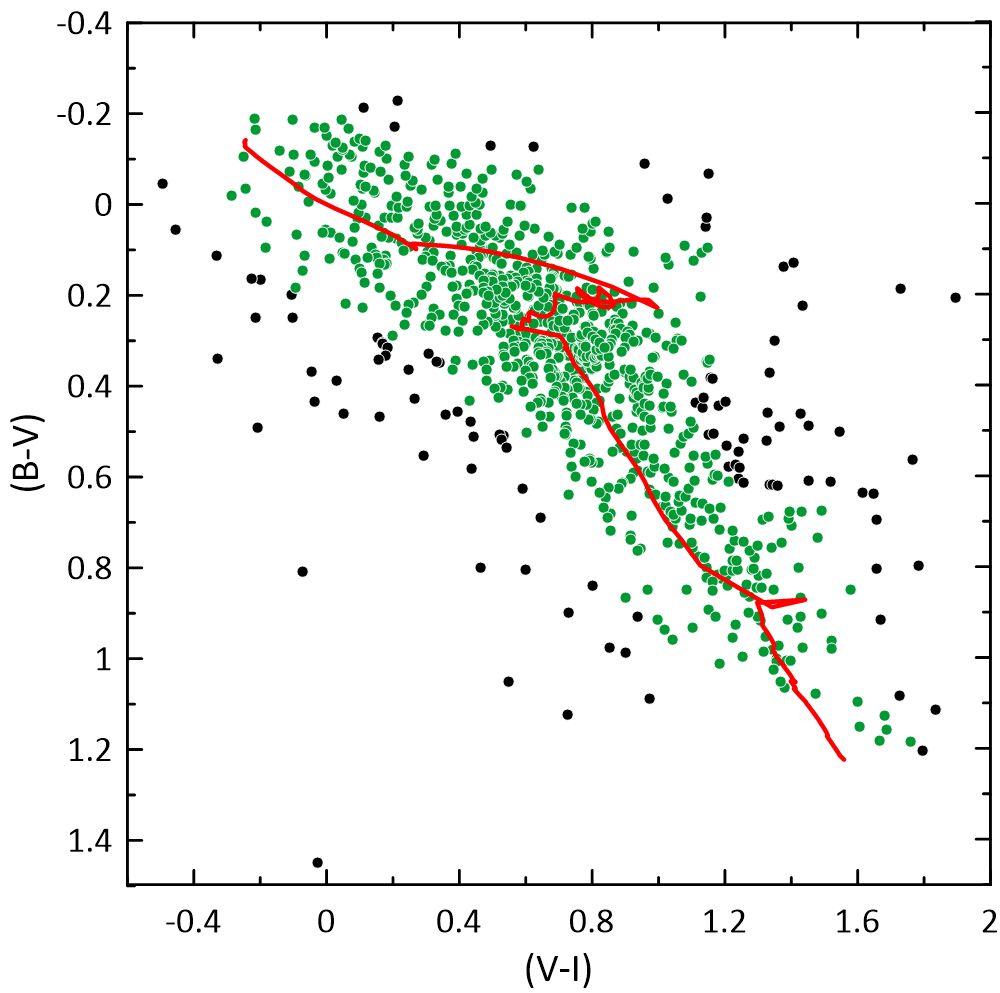}
\caption{ Color-color diagram of the star clusters in NGC 628. The red solid line corresponds to selected SSP model. The green and black dots represent the clusters  those with $\leq$1$\sigma$ and $>$1$\sigma$ standard deviation from the  SSP model, respectively.} 
\label{F:CCSSP}
\end{center}
\end{figure}

Comparing color-color diagrams with a Simple Stellar Population (SSP) is important for accurate determination of both cluster age and mass. Also, this diagram can help to distinguish the clusters from stars. Here, we estimate the age and mass of each cluster by comparing the population synthesis models of \cite{2003MNRAS.344..1000B}. The color$-$color diagram of the star clusters for NGC 628 are plotted in Fig. \ref{F:CCSSP}. This diagram is constructed by using an SSP model of \cite{2003MNRAS.344..1000B} with Kroupa initial mass function (IMF) for solar metallicity (Z=0.02) using mean extinction as E(B-V)=0.15 mag. But in fact, we estimated extinctions  individually for each cluster by moving SSP according to the reddening line in two directions (horizontally and vertically) with different metallicities by performing  $\chi^2$ fits. For the extinction coefficients between filters, we used Table 14 from  \cite{2005PASP..117.1049S}. Then, we estimated the age for each cluster by comparing the observed colors with the predictions from SSP models.

In order to eliminate the scattered points 1$\sigma$ (standard deviation) from the SSP model was taken into account. As a result, a total of 106 clusters were eliminated. All analyzes were performed for the remaining 864 clusters. The presence of clusters above the SSP deviating from the models may be the result of higher extinctions of these clusters. Similarly, the clusters below the model could be misclassified objects located in starry location as mentioned in \cite{2010ApJ...719..966C}. In our study, the percentage of misclassified objects is only $\approx{2\%}$ in total indicating our cluster selection criteria are reliable.

\begin{figure}
\begin{center}
\includegraphics[angle=0,scale=0.38]{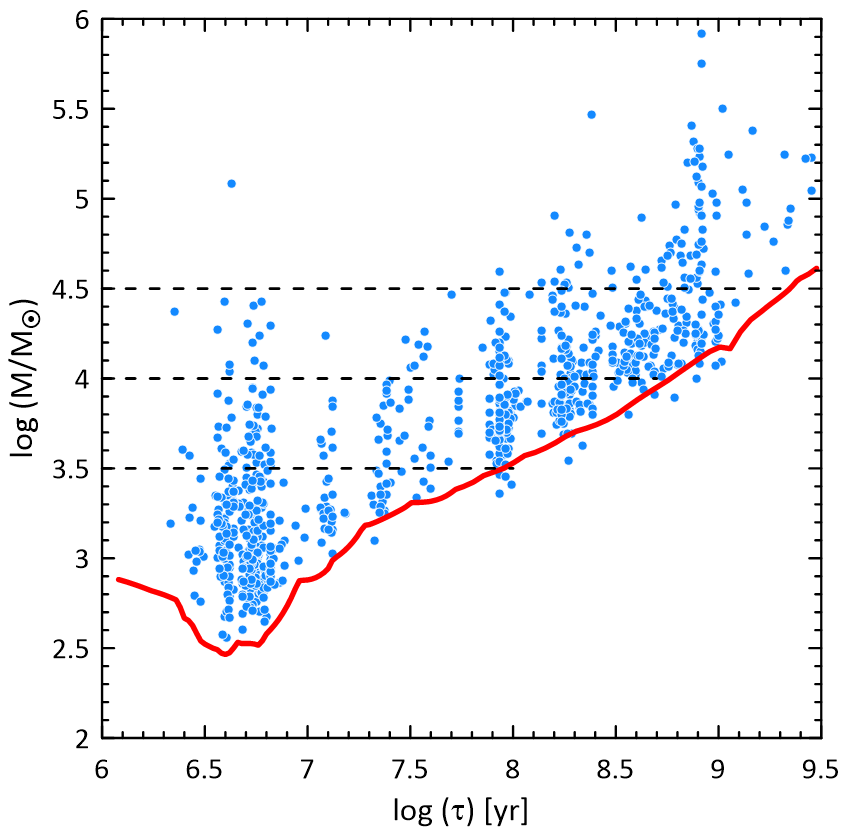}
\caption{Mass-age diagram of the star clusters ($N \approx 864$) in NGC 628. The red solid line shows the approximate magnitude limit of $M_{V}$=-6.2 mag. The dashed lines show the two different intervals of mass and represent the population well due to its covering a wide age interval.}
\label{F:AgeMass}
\end{center} 
\end{figure}

The mass and age distributions for a population of star clusters provide information about cluster formation and disruption in a galaxy. Accordingly the mass-age distribution of the star clusters derived from SSP model is given in Fig. \ref{F:AgeMass}. The approximate completeness magnitude limit of $M_{V}$=-6.2 mag (m$_{F555W}$ $\simeq$ 24 mag), is shown with the solid red line as it ages from 6 Myr to 9.5 Gyr. Although we choose a magnitude limit as m$_{F555W}$ <25 mag for the cluster selection at the beginning, we saw the clusters with $>$1$\sigma$ standard deviation from the SSP model in Fig. \ref{F:CCSSP} have a magnitude range between 24 and 25 mag. So, we decided to use $M_{V}$=-6.2 mag as an approximate 90$\%$ completeness limit for the mass, age and luminosity distributions. In addition, this limit value is compatible with the cluster mass-age distribution in Fig.  \ref{F:AgeMass}  

As seen in Fig. \ref{F:AgeMass}, the most clusters in NGC 628 disrupt over the time between the ages 7$<$ log $\tau$ $<$8 yr. The reason for this could be that the less massive and less concentrated clusters begin to fade over given age range by stellar evolution and tidal interactions and disappear at older ages. Therefore, the distribution in this figure is compatible with the general evolution of the clusters in a galaxy so that the massive clusters are seen at older ages than for low mass clusters. The minimum mass limit for clusters at younger ages is around log($M/M_{\odot}$)=2.6 and the maximum mass limit reaches log($M/M_{\odot}$)=5.5 at log($\tau$)=9.0 yr in NGC 628. \cite{2017ApJ...841..131A} and \cite{2015ApJ...815...93G} found similar mass-age distributions for the clusters in NGC 628 although the number of the clusters are not the same with ours. 

\begin{figure*}
\begin{center}
\includegraphics[angle=0,scale=0.33]{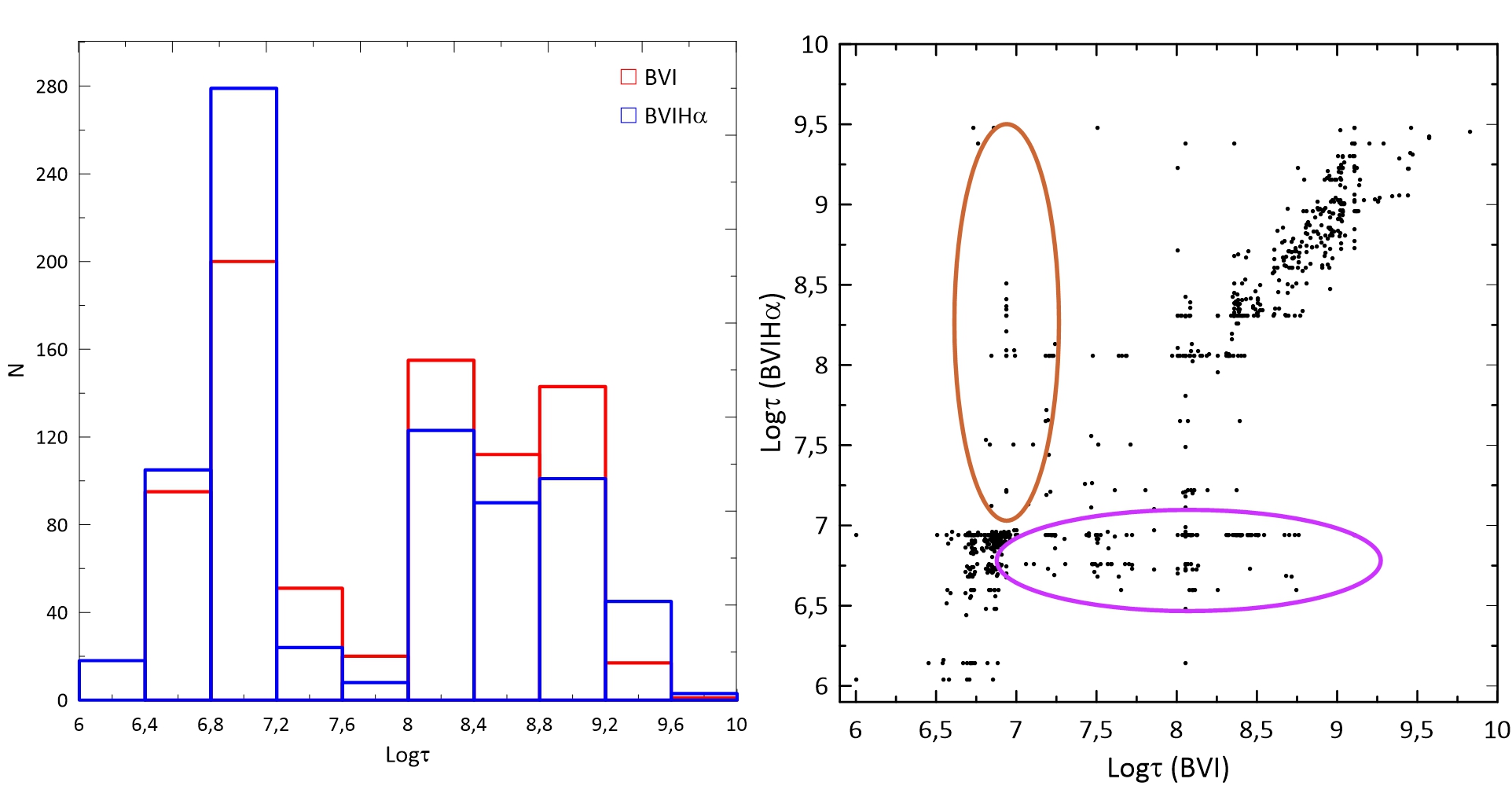}
\caption{Comparison of the cluster ages obtained using BVI and BVIH$_{\alpha}$ filter combinations (includes clusters in the Center and West fields). In the left panel, red and blue lines represent number of the clusters in different ages derived using BVI and BVIH$_{\alpha}$, respectively. The right panel shows the change in age of the same clusters as calculated using BVI and BVIH$_{\alpha}$. Clusters in brown and purple ellipses are described in the text.} 
\label{F:bviha}
\end{center}
\end{figure*}

In our study, we constructed another color-color diagram to determine the cluster mass and age by including {\it HST}/ACS F658N (H$_{\alpha}$) filter as (F435W-F555W vs F658N-F814W). Because this filter effects the age and mass of the cluster due to the nebular and continuum emissions and shows the age-extinction degeneracy. For this galaxy, F658N filter is available for the center and west fields. From the constructed color-color diagram with F658N filter, we calculated ages for the clusters assuming the same metallicity, SSP model and extinction. Comparison of ages determined from these filter combinations is shown in Fig. \ref{F:bviha}. The left panel shows the increase in the number of clusters $\tau$ <$10^{7.2}$ yr when the F658N filter is included. Likewise, It is represented by the purple ellipse, where with the BVI filters some clusters are older in age, but with the BVIH$_{\alpha}$ filter combination the same clusters are relatively younger. We think that the presence of H$_{\alpha}$ emission indicates that these clusters are  moderately reddened young clusters. As highlighted by \cite{2016ApJ...824...71C},  it is more reliable to include measurements of H$_{\alpha}$ or other appropriate narrowband filter results when estimating the age of star clusters.
Indeed, we found that these reddened clusters are located along the spiral arms and HII regions in this galaxy. Conversely, a few clusters (clusters in the brown ellipse in the right panel of Fig. \ref{F:bviha}) have older ages when F658N filter is included. This could be due to lack of nebular emission in these regions. In the study of \cite{2016ApJ...824...71C} the same filter (F658N) effects were reported for the star clusters in M51.

\subsection{Comparison of Mass and Age Distributions with Disruption Models}

Star clusters are born in large molecular clouds and become disrupted due to physical processes throughout their lives. The mass and age distribution for a population of clusters in a galaxy are given with $\psi(M) \propto dN/dM$ and $\chi(\tau) \propto dN/d\tau$, respectively. \cite{2009ApJ...704..453F} derive formulae for bivariate mass–age distribution $g(M,\tau) \propto \psi(M)\chi(\tau)$, for three different models for the disruption of clusters. These models are defined as sudden mass-dependent disruption (Model 1), gradual mass-dependent disruption (Model 2) and gradual mass-independent disruption (Model 3). \cite{2009ApJ...704..453F} also present analytical formulae for the mean mass and age distributions ($\bar{g}(M)$ and $\bar{g}(\tau)$) in their Appendix B for these three models. Here, we constructed $\bar{g}(M)$ and $\bar{g}(\tau)$ for the clusters in Fig. \ref{F:AgeMass} and compare those to the analytical predictions for Model 1, Model 2 and Model 3.

We divided the mass distribution into two age intervals as <10 Myr and $\geq$100 Myr and the age distribution is divided into two different intervals of mass as $\log(M/M_{\odot})=3.5-4.0$ and $\log(M/M_{\odot})=4.0-4.5$. The selected mass ranges are well representative of the population as they cover a wide age range as seen in Fig. \ref{F:AgeMass}. We also checked the mass and age distribution of star clusters if they are sensitive to different bin widths. We found that it is not sensitive and therefore, we chose a bin width of $\triangle\log(M)=0.2$ for $\bar{g}(M)$ and a bin width of $\triangle\log(\tau)=0.4$ for $\bar{g}(\tau)$. 
 
\begin{figure}
\begin{center}
\includegraphics[angle=0,scale=0.4]{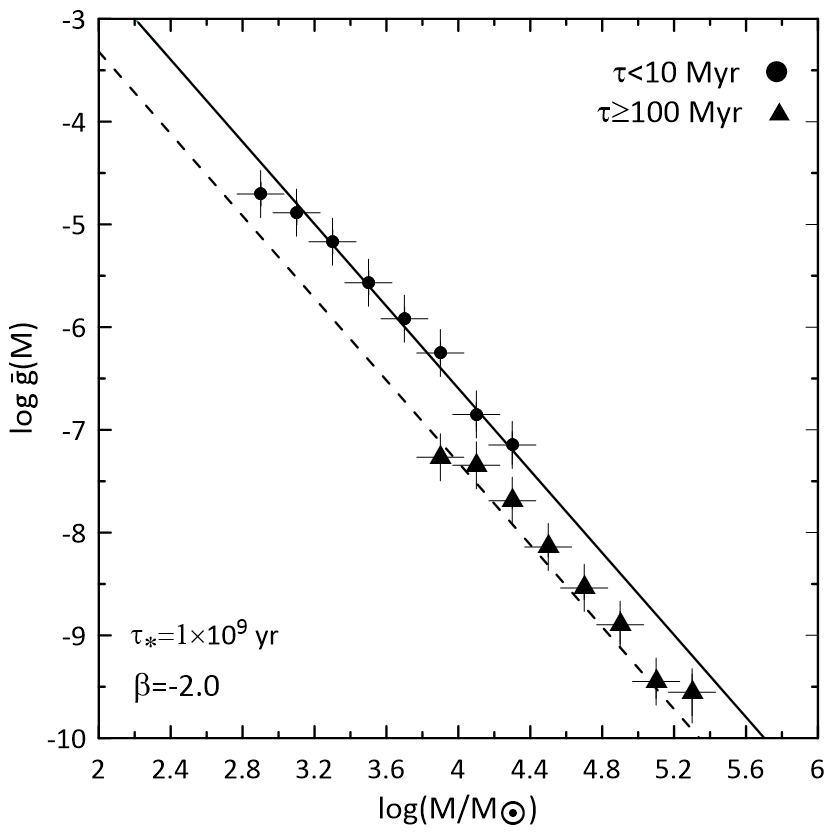}
\includegraphics[angle=0,scale=0.4]{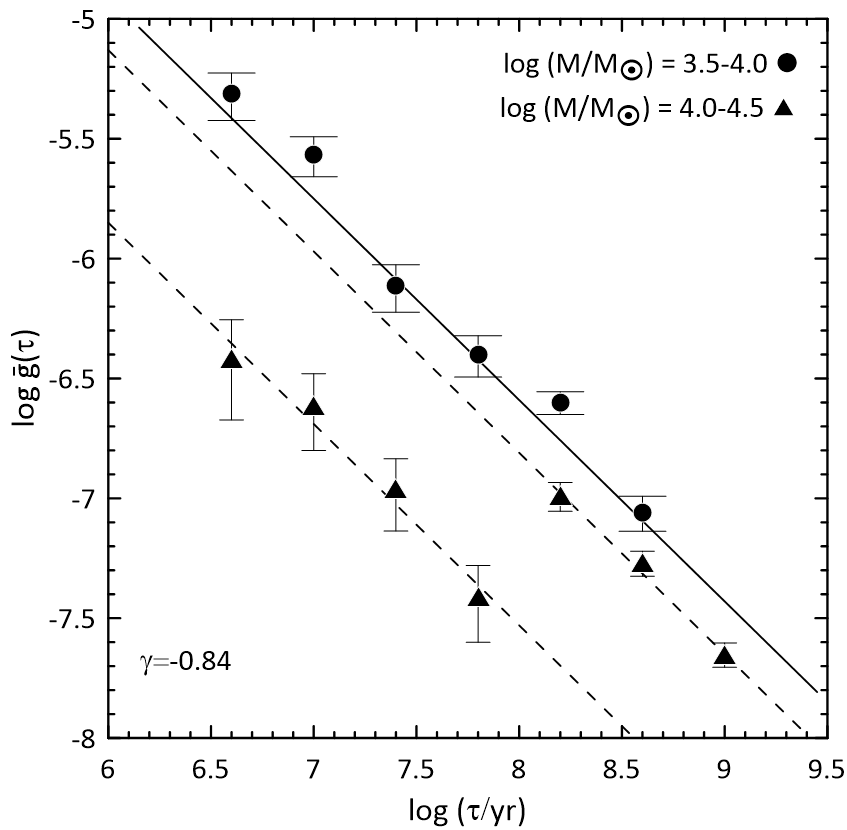}
\caption{ The comparison of the observed and predicted mass function (upper) for the clusters in NGC 628 for Model 3. Filled circles and triangles represent the observed mass distribution averaged over the indicated age intervals. The solid and dashed lines were computed for the younger and older clusters with $\beta=-2$ for Model 3. Age distribution (lower) of the clusters in different mass intervals for Model 3. The solid and dashed lines were computed  over the indicated mass intervals with $\gamma=-0.84$ for Model 3.} 
\label{F:MA_Model3}
\end{center}
\end{figure}

We noticed that the observed $\bar{g}(M)$ of clusters declines monotonically in two different age intervals with no obvious breaks as seen in the upper panel of Fig. \ref{F:MA_Model3}. Also $\bar{g}(M)$ follows a power-law model without flattening at lower masses. Similarly, observed $\bar{g}(\tau)$ distribution of star clusters decline with a power-law in age for the given two mass intervals (lower panel in Fig. \ref{F:MA_Model3}). These observed mass and age distributions do not fit adequately with Model 1 and Model 2. Because the expected flattening at lower masses from the models do not appear in the observed $\bar{g}(M)$ distribution and also the observed $\bar{g}(\tau)$ distribution does not remain flat for a long time but decline with a power-law. 

Therefore, we applied a gradual mass-independent disruption model (Model 3) to the young and old cluster populations in the galaxy. In this model, the mass function is independent of age and also the age function is independent of mass. The mass and age distributions are represented by a gradual decrease with a power-law. The observed cluster distributions with the applied Model 3 is given in Fig. \ref{F:MA_Model3}. The given mass and age distributions in upper and lower panel in Fig. \ref{F:MA_Model3} are compatible in all age and mass intervals. The power-law indices were found to be $\beta$=-2.00$\pm$0.12 for mass function and $\gamma$=-0.84$\pm$0.20 for age distribution. As seen in the upper panel, the observed mass function has the same power-law index for different ages but amplitudes have not. The differences in amplitudes between the young (log$\tau$ <7.0) and old (log$\tau$ >8.0) cluster populations could be interpret as the variation of the cluster formation rate in different age intervals \citep{2010ApJ...711.1263C}. It is expected that the cluster formation rate for the same mass intervals will decrease as time passes. Here, we show that there is not a similar trend for the massive clusters ($\log(M/M_{\odot})$ =4.0-4.5)) due to a jump in $\bar{g}(\tau)$ around log$(\tau/yr) \sim$8.2 as seen in the bottom panel of Fig. \ref{F:MA_Model3}. This can be explained by the fact that the cluster formation rate of the massive and older star clusters was previously higher. Which could be interpreted as gas falling towards the center during the early evolutionary stages of the galaxy, triggering the formation of massive clusters \citep{2011AJ....142...16Z}. Considering that star clusters are associated with star forming regions, this interpretation is also compatible with recent study of \cite{2022MNRAS.516.2171U} where they suggest that  the recently formed star forming regions are less massive than the older star forming regions in NGC 628.

The gradual mass-independent model is a best fit for all age and mass intervals with $g(M,\tau)$ $\propto$ $M^{-2.0}$ $\tau^{-0.84}$ for star clusters in NGC 628. \cite{2009ApJ...704..453F} and \cite{2010ApJ...711.1263C} compared Model 1, Model 2 and Model 3 for star clusters in Antennae and LMC. \cite{2009ApJ...704..453F} found that the mass-independent disruption model to be the most successful model and has the simple form $g(M,\tau)$ $\propto$ $M^{-2}$ $\tau^{-1}$ for massive young clusters in the Antennae. \cite{2010ApJ...711.1263C} also reported that Model 3 provides a good description for LMC cluster populations, with $g(M,\tau)$ $\propto$ $M^{-1.8}$ $\tau^{-0.8}$. Although the slopes found for the mass and age distributions vary from galaxy to galaxy as seen in the given studies, the slopes found for NGC 628 star clusters are consistent within the error limits. 

\subsection{Luminosity Function of Clusters}

\begin{figure}
\begin{center}
\includegraphics[angle=0,scale=0.43]{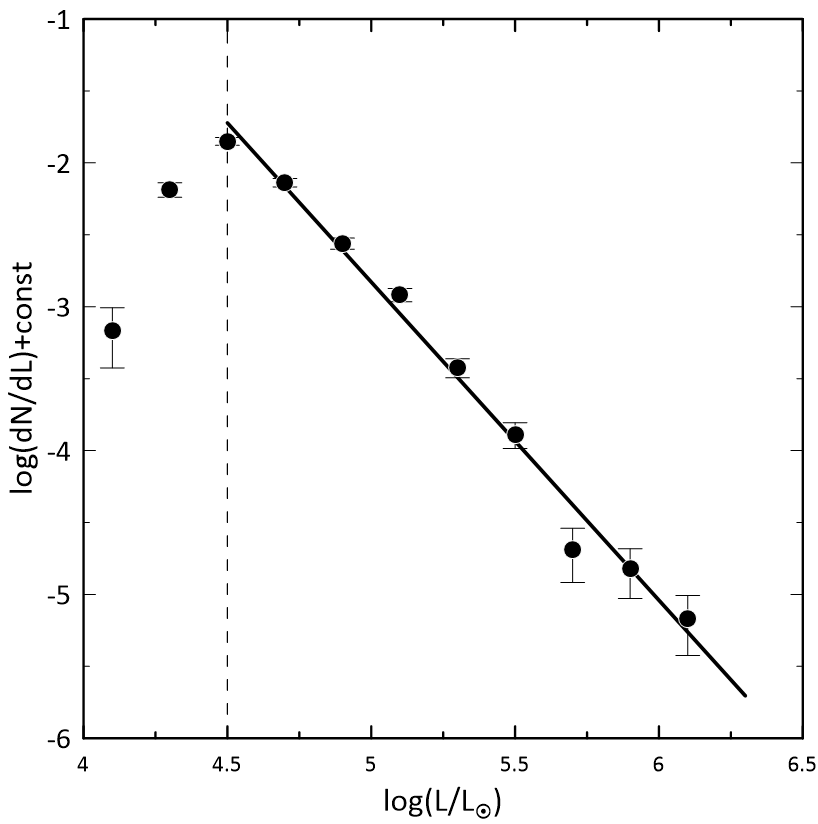}
\caption{Luminosity function of clusters in the NGC 628 in the V band. The magnitudes have been corrected for extinction. The dashed line represents luminosity limit of clusters corresponding to the magnitude of $M_{V}$=-6.2.} 
\label{F:LF}
\end{center}
\end{figure}

We have also derived the luminosity function (LF) of the NGC 628 clusters since LF derived directly from the observations and not affected by age or mass determinations. In Fig. \ref{F:LF} we show extinction-corrected V-band luminosity function for our star clusters. The luminosities include our size-dependent aperture corrections and have been corrected for the extinction we found. We used constant bin sizes with $\triangle\log(L/L_{\odot})=0.2$. This LF can be approximated by a single power-law, $\phi(L) \propto L^{\alpha}$ with $\alpha=-2.19\pm0.15$. The slope of LF is identical within the uncertainties to the cluster mass function (MF) ($\alpha=-2.00\pm0.12$) as Fig. \ref{F:MA_Model3} shows that the galaxy undergoes a mass-independent evolution \citep{2010ApJ...711.1263C}. \cite{2017ApJ...841..131A} obtained the LF of the whole cluster population of NGC 628 in the five standard LEGUS bands. They build the LFs by using different filters for the sources that have been visually classified as class 1, 2, or 3 and the slopes found are close to an index of $-$2. \cite{2022MNRAS.509..180L} present the LF of globular star clusters in this galaxy with {\it HST} observations. They choose the extinction-corrected magnitudes in the F814W filter to construct the LF and found a turning point in LF. Although they did not give a slope for LF, a smooth decrease towards the fainter magnitudes in their LF was seen.

\section{Results and Discussions}
\label{sec:discussion}

In this study, we investigated the identification of donor stars of XRBs and their relationship to the ages of nearby compact star clusters in order to determine different classes of them. We noticed that only 42 of detected XRBs fall within the {\it HST} field of view and 18 of them were associated with star clusters in NGC 628. The positions of XRBs and the properties of associated clusters are given in Table \ref{T:x-optic}. {\it HST}/ACS true color images of corrected position of these XRBs are given in Fig. \ref{F:3color}.
In this figure, the position of X55 is not shown as it falls to the edge of the chip. The bright X-ray source ULX-1 is also shown in the Figure, but it can not be associated with any of star clusters. Therefore, it will be discussed separately.


\subsection{The Connection Between XRBs and Star Clusters}
\label{sec:5.1}

To figure out whether XRBs are bound to cluster candidates, we derive a chance alignment probability. For this, we followed similar technique given in the recent study of \cite{2022AJ....164....7P}. In this calculation, 18 XRBs and 864 cluster candidates in the 3.2$\arcmin$ $\times$ 3.3$\arcmin$ field of view of {\it HST} are taken into account and then, a cumulative distribution of the number of sources as a function of the apparent magnitude of F555W was constructed. Assuming a random uniform Poisson distribution of unrelated XRBs and clusters across the given area, the probability that a XRB is not related to the cluster is given by P$_{un}$($\sigma$, $\Delta$)=1 $-$ exp(-$\pi\sigma\Delta^{2}$), where $\sigma$ is the surface density determined by dividing the result of the cumulative distribution by the area. $\Delta$ is the projected separation from XRB (in arcsec) to the nearest cluster. These parameters and the value of being related to the cluster, 1 $-$ P$_{un}$($\sigma$, $\Delta$),  are given in Table \ref{T:prob}. As assumed in \cite{2022AJ....164....7P}, if the probability values are between 60\% and 90\%, we consider these XRBs to be related with clusters, and to be more convincing if this probability is >90\% . As noted from Table \ref{T:prob}, the chance alignment probability of X5 and X75 are high due to the large $\Delta$ and $\sigma$ values.

\begin{figure*}
\begin{center}
\includegraphics[angle=0,scale=0.25]{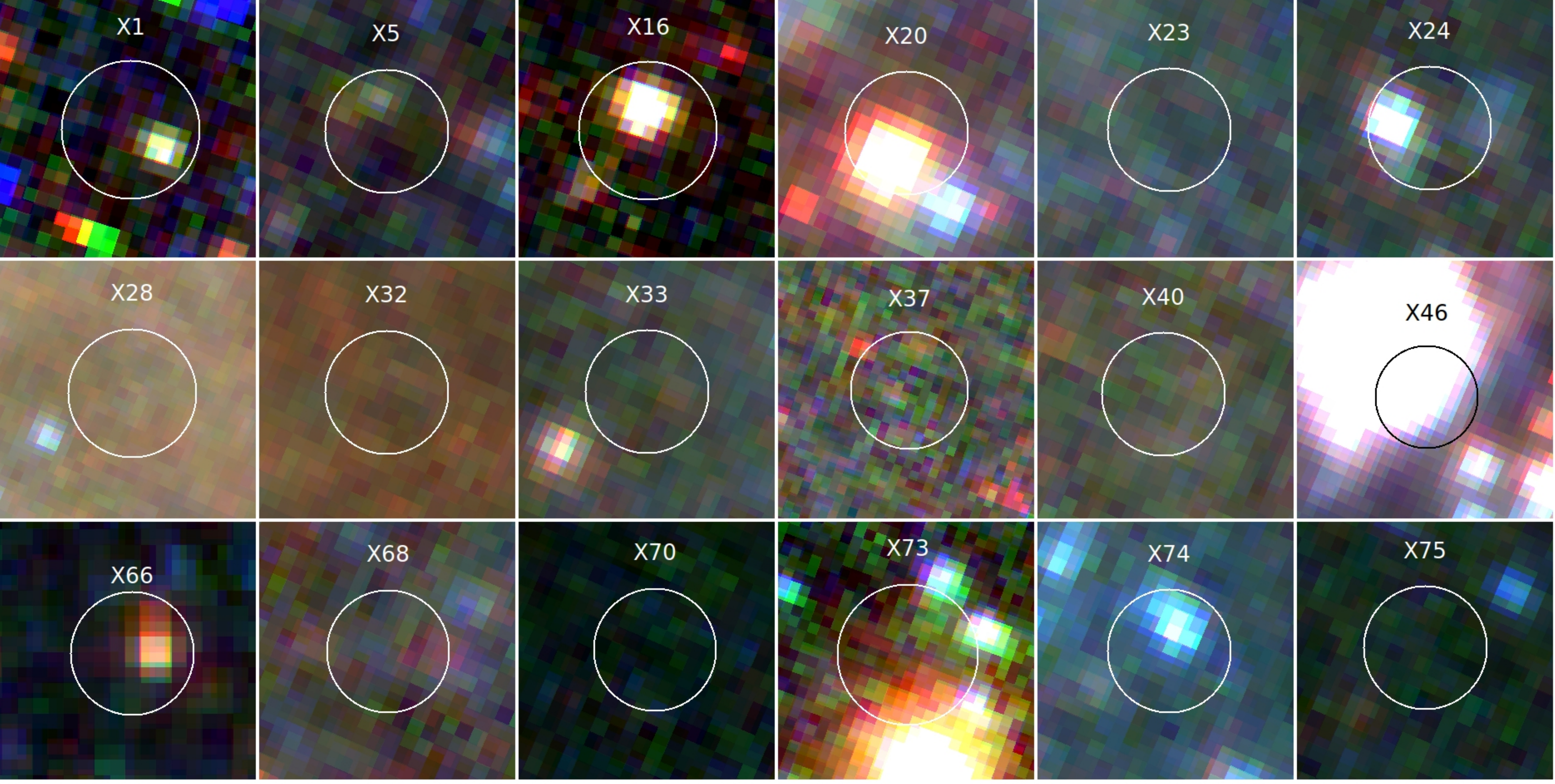}
\caption{{\it HST}/ACS true color images of corrected positions of ULX X-1 and X-ray sources which are associated with star clusters within their 200 pc.} The white and black circles show the astrometric error radii. Blue, green and red colors represent {\it HST}/ACS F435W, F555W and F814W filters, respectively. North is up.
\label{F:3color}
\end{center}
\end{figure*}

We identified that 6 out of 18 XRBs have unique optical candidates, that may act as the donor star, within the 2$\sigma$ positional uncertainty. Therefore, X5, X16, X20, X24, X66 and X74 could be classified as HMXBs. Their optical candidates were already shown in Fig. \ref{F:3color}. In Fig. \ref{F:cmd}, the relevant mass tracks are given from Padova Isochrones for their donor stars (except for X5) with having masses in the range $\sim 10-15 M_{\odot}$. Because X5 has a very faint optical candidate in the {\it HST}/ACS F435W filter; we used the other CMD (F555W versus F555W-F814W) to estimate its mass as $\sim 7 M_{\odot}$. Considering the mass range of IMXB (3$-$8 M$_{\odot}$) we can, therefore, classify X5 as an IMXB. For the remaining five sources their spectral colors (-0.1<$(B-V)_{0}$<1.2) and absolute magnitudes (-5.3 < M$_{V}$ < -7.2) indicate that optical counterparts might be in the giant phases of their lives. HMXBs often occur in young star clusters and their bright optical counterparts allow us to identify them. Many of them are found close (< 200 pc), but not within any young cluster. In the present case, the six sources mentioned are all associated with young clusters within 200 pc and with ages $\tau$ < 10$^{8.5}$ yr. In fact, although these HMXBs were born in young clusters, they might be kicked out from their hosts. The ejection scenarios are basically explained by three different mechanisms \citep{2002ApJ...577..710Z, 2004MNRAS.348L..28K, 2011ApJ...741...86R, 2012ApJ...755...49G}. Firstly, these systems may have been expelled from the cluster as a result of a dynamical interaction with other stars in the dense cluster core. Secondly, the system may have been ejected from the host at the end of an asymmetric supernova explosion within the cluster. Thirdly, as the host cluster quickly dissolves after XRBs formed.

Although it may be difficult to say anything about for 6 XRBs in which way they are kicked out, we can impose constraints on their ejection velocities. Considering the displacements between XRBs and star clusters (given in Table \ref{T:x-optic}), we determined the ejection velocities of six sources and found their lower limits as $\sim$1-10 km s$^{-1}$. At this point we need a cautionary note that, our velocity calculation uses the entire cluster life time, therefore, special considerations need to be taken; for example, \cite{2011ApJ...741...86R} assumed that ejection occurred about 1Myr ago to obtain an upper velocity limit. 

\begin{figure}
\begin{center}
\includegraphics[angle=0,scale=0.4]{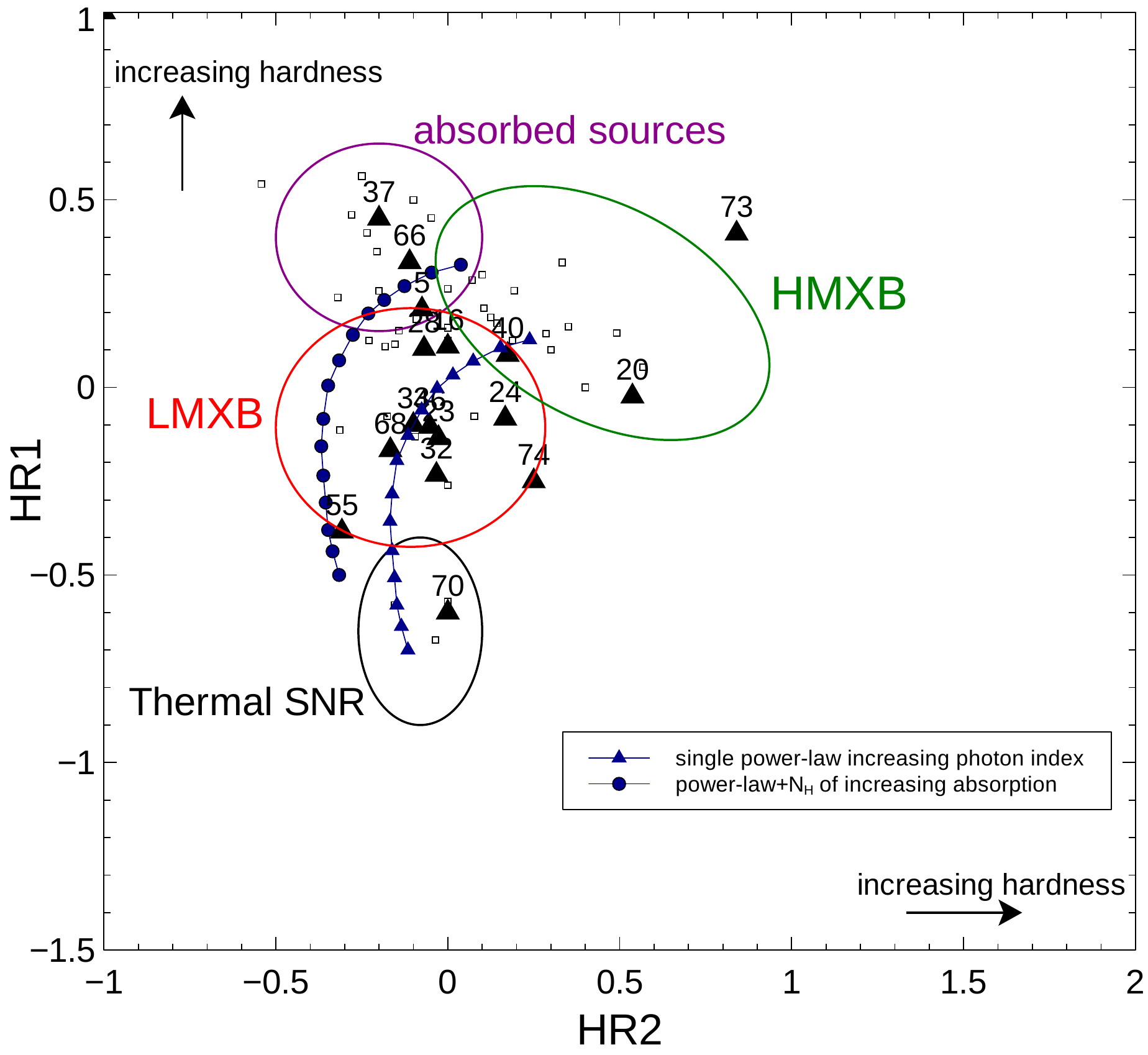}
\caption{X-ray color-color diagram of all detected X-ray sources (open squares) within the D$_{25}$ region in NGC 628. The sources that may be associated with clusters are shown with filled triangles and X-ray ID are labeled. Blue triangles and circles represent simple power-law spectral model with $\Gamma$ increasing from 0.7 to 3.0 and adding galactic absorption to power-law model, respectively. Red, green, black and magenta circles were accepted from LXMB, HMXB, thermal SNR and absorbed sources classification in \protect\cite{2003ApJ...595..719P}, respectively.}
\label{F:2color}
\end{center}
\end{figure}

The following nine X-ray sources (X23, X28, X32, X33, X40, X68, X70, X73 and X75) are classified as LMXBs, because they have no optical counterparts and are located close to the clusters with ages, $\tau$ > 10$^{8.4}$ yr. Also, when the {\it HST}/ACS F658N image was examined, these sources and associated clusters were not found in the immediate vicinities of any HII regions. The X-ray color-colors of the five of these sources (X23, X28, X32, X33, X40 and X68) are consistent with their classification as LMXBs (see Fig. \ref{F:2color}). In this Figure, it also appears that the colors of most of these sources agree with the theoretical track of the power-law model (1.5 < $\Gamma$ < 2.1). It is discussed in the literature that the accretion spectra of low-mass neutron star binaries are adequately fitted by the power-law model with a photon index of $\Gamma$ $\sim$ 2.0 \citep{2003ApJ...595..719P, 2011ApJ...741...86R}. On the other hand, X70 from the remaining 3 sources is located in a soft X-ray source region containing thermal SNRs in the color-color diagram, however, its transit nature we have identified does not allow us to classify it is an SNR. Considering its luminosity ($\approx$10$^{37}$ erg s$^{-1}$ ), it could be classified as soft X-ray transient, a subclass of LMXBs \citep{2002A&A...393..205E, 2020AdSpR..66.1004H}. For the other two sources, X73 and X75, it is not possible to classify them according to their positions on color-color diagram since their count rates in the soft energy range cannot be measured. When their 3-$\sigma$ upper limit values were taken, these sources do not fall into the appropriate range in Fig. \ref{F:2color}.

We noticed that the sources X46 and X55 given in Table \ref{T:x-optic} and the associated clusters are located within HII regions. Ages of these clusters were calculated as < 10$^{7}$ yr. However, these sources need further elaborations: Both of them are classified as LMXB, therefore they need to be associated with old clusters. We think that the presence of H$\alpha$ emission indicates that these clusters are  moderately reddened young clusters. As highlighted by \cite{2016ApJ...824...71C}, we consider it to be more reliable to include measurements of H$\alpha$ or other appropriate narrow band filters when estimating the age of star clusters. Both sources were classified as LMXBs since their optical counterparts could not be determined, but there is some ambiguities for this interpretation because the astrometric error circle of X46 matches the bright emission region as seen in Fig. \ref{F:3color}. On the other hand, half of the error circle of X55 corresponds to the chip gap in the {\it HST} image.  The young age of the nearby clusters allow us to comment that maybe these sources (X46 and X55) have optical counterparts, but we could not identify them in the {\it HST} images. But the same interpretation is not possible for X37. This source are classified as LMXB and is not located in HII regions. When carefully examined, it is found that a young star cluster is quite close to this XRB. Although there is a low probability of chance alignment, this source may or may not be associated with the cluster due to the age incompatibility.

\begin{figure}
\begin{center}
\includegraphics[angle=0,scale=0.5]{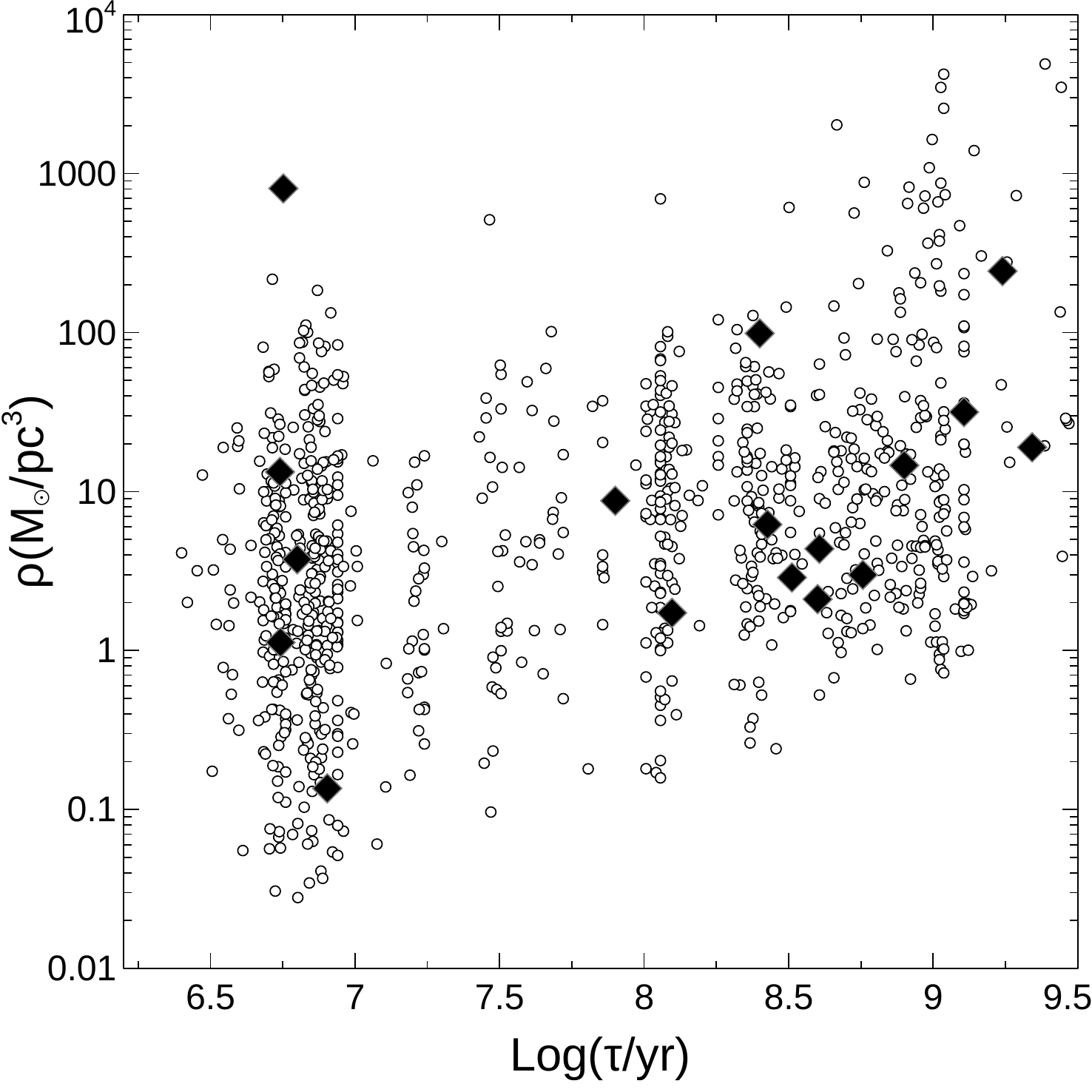}
\caption{Age and density distribution of star clusters in NGC 628. The black unfilled dots show 864 clusters and black filled diamonds represent host clusters of XRBs. } 
\label{F:dist}
\end{center}
\end{figure}

To see if there are notable trends for LMXBs and HMXBs in the age and density distribution of star clusters, we show their distribution in Fig. \ref{F:dist}. We notice that the clusters related to LMXBs are relatively dense. This indicates that dynamical interactions in clusters play an important role in the formation of LMXBs. However, the same trend is not seen for HMXBs. Our findings show similar distributions found by \cite{2019ApJ...871..122J} for Antenna galaxies.

As we mentioned, 320 compact star clusters were cataloged by \cite{2017ApJ...841...92R} in the east and center fields of NGC 628, using {\it HST}/ACS and WFC3 images. Although some of our clusters are outside of these two fields, 70$\%$ of their star clusters match in the position of ours. On the other hand, locations of 42 XRBs we identified within the same {\it HST} fields were compared with their cluster sample. We found that only 6 of them (X5, X21, X23, X45, X58 and X66) were associated with the clusters within distance of 200 pc of radius.

Although the source X1 doesn't match any cluster being the most luminous source in our list, we also took a closer look at the X1. Previously, X1 was already classified as a variable ULX by \cite{2005ApJ...630..228K} using 2001 {\it Chandra} and 2002 {\it XMM-Newton} observations. To confirm this, we analyzed all {\it Chandra} observations between 2001 and 2018 and see that X1 is indeed a persistent and highly variable (V$_{f}$ $\sim$ 50) one. In addition, after the astrometric correction, we identified that X1 has a unique optical counterpart within the astrometric error circle (Fig. \ref{F:3color}). Assuming the optical emission is dominated from the donor star of X1 it is spectral class would be determined G type supergiant by using color and absolute magnitude tables of \cite{1982lbg6.conf.....A}. The age and mass of donor star were also calculated as $\sim$16 Myr and $\sim$13 M$_{\odot}$ from CMDs we constructed. The presence of its optical counterpart and the mass of the donor star indicate that X1 can be classified as a HMXB.

\subsection{X-ray Variability of XRBs}

\begin{figure}
\begin{center}
\includegraphics[angle=0,scale=0.45]{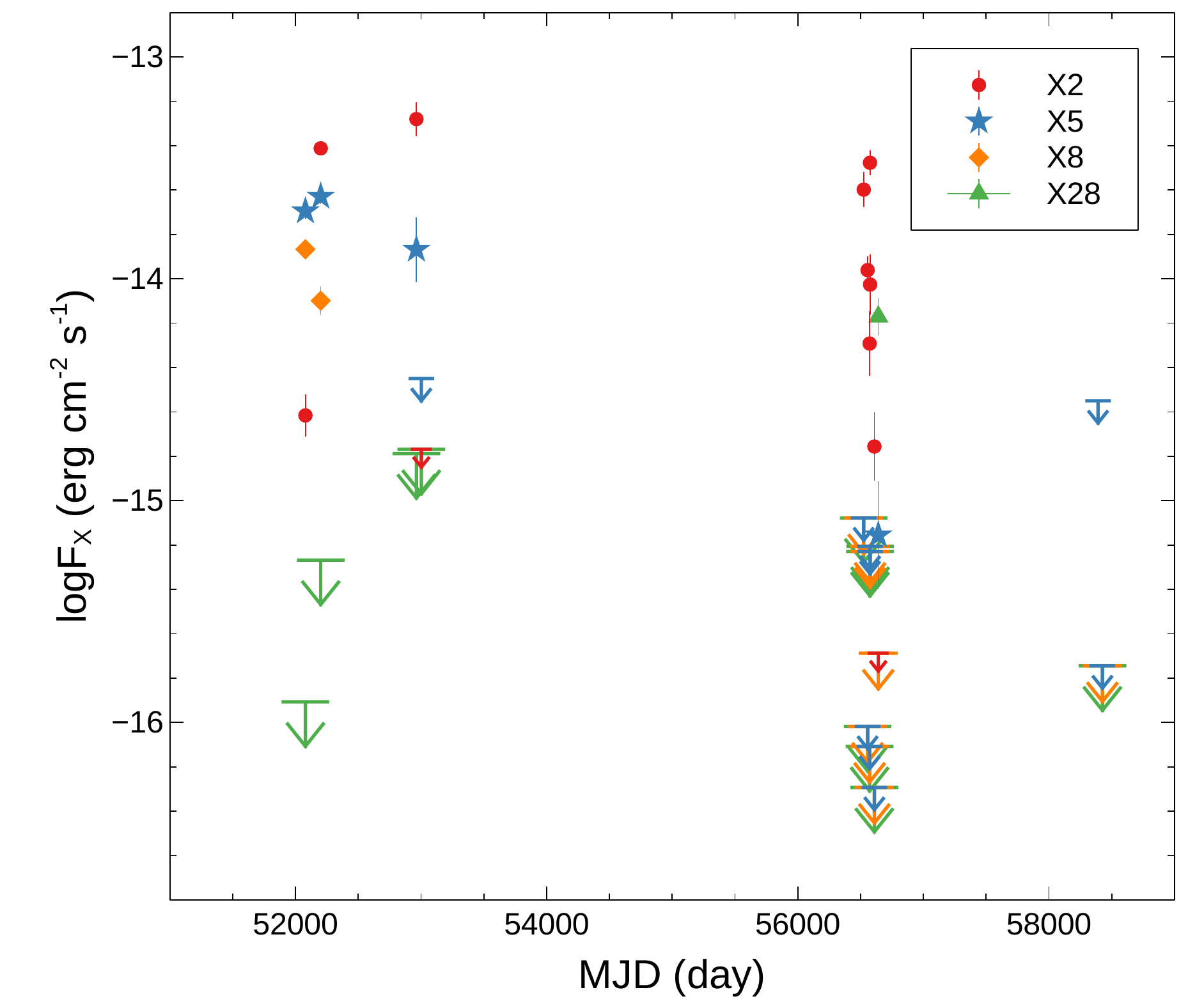}
\caption{The long-term light curves of 4 X-ray sources with the highest variability in NGC 628. The fluxes were calculated in the 0.3$-$8 keV energy range.}
\label{F:lc}
\end{center}
\end{figure}

The X-ray variability of 42 XRBs in NGC 628 was investigated to understand their X-ray properties. Temporal features of these XRBs obtained from the analyses of 13 data sets over 17 years are given in Table \ref{T:x-optic}. The long-term light curves of some XRBs which exhibit high variability are given in Fig. \ref{F:lc}. No clear correlation is observed between the classification of these sources (as LMXB or HMXB) and their X-ray variabilities (see columns 4 and 5 in Table \ref{T:x-optic}).

The Galactic XRB population is dominated by LMXBs with compact object neutron stars or black holes. Most of these systems do not continuously absorb matter via accretion, so they classify as transient sources. They occupy very low luminosity states in quiescence, but occasionally show outbursts with high levels of accretion onto the compact object that typically  last for weeks or months time-scales \citep{ 2013A&A...556A...9Z,2022arXiv220610053B}.
As seen in Table \ref{T:x-optic}, according to the variability factor, 26 of the sources that we defined as transient or transient candidates are LMXBs. Unfortunately, due to insufficient data points,  the duration of outburst  for these transient sources is unclear. Therefore, it is difficult to interpret whether they are compatible with the duration of the Galactic LMXBs outbursts.

On the other hand, HMXBs are divided into two subgroups according to whether the primary star is a Be star (Be/XRB) or a supergiant (SG/XRB). Majority of the known Galactic HMXBs are Be/XRBs where accretion on the compact object occurs via equatorial disk formation by the stellar wind. 
Most Be/XRBs have a transient nature, but the existence of persistent Be/XRBs have also been reported.  \citep{2011Ap&SS.332....1R,2013A&A...553A.103F,2022A&A...662A..22H}. In our present study, 11 of the 42 X-ray sources are classified as transient HMXBs with a variability factor, V$_{f}$> 10 (see Table \ref{T:x-optic}). 5 of these HMXBs (X5, X56, X58, X65 and X74) have donor stars with spectral type B according to their color index (B-V) and M$_{V}$ values. These sources have high variability (V$_{f}$>50) and their transient nature is reminiscent of Be/XRBs.

\subsection{Optical Emission of XRBs}

In the present study, we identified the possible optical counterparts of the XRBs in NGC 628 with precise astrometric calculations based on {\it Chandra} and {\it HST} data. The problem with identifying the donor star in the optical emission of XRBs is that the flux contribution from the accretion disk, either directly or irradiated, is mostly unknown. Using {\it HST} UV/optical observations, we investigated the color and SED (Spectral Energy Distribution) features of the optical counterparts to analyze their possible contribution to the observed optical emission.

We try to constrain the nature of counterparts by fitting SED with a blackbody or a power-law spectrum. To obtain a more reliable fitting, we considered sources detected in more than three filters. Only five counterparts (X1, X20, X24, X34 and X58) meet this condition, but no acceptable model has been determined for any of them. Also, if the optical emission is dominated by the accretion disk, then the optical variability should be high \citep{2017ARA&A..55..303K,2022MNRAS.515.3632A}. However, this is difficult to interpret since simultaneous optical and X-ray observations of the sources are not yet available.

\section{Summary and Conclusions}
\label{sec:summary}

In this study, the association of 42 XRBs of NGC 628 with compact star clusters in the {\it HST}/ACS fields were investigated. For this, we identified the relevant compact star clusters and investigated their mass, age and luminosity distributions. The disruption models of clusters are also discussed. Our main findings can be summarized as follows:

\begin{enumerate}
    \item Using the cluster selection criteria ($m_{V}$< 25 ($M_{V}\approx$ $-6.2$); CI > 2.3; FWHM > 0.2 pixels, and no other object within a radius of 5 pixels), 864 star clusters in NGC 628 were determined within < 1$\sigma$ standard deviation from SSP model tracks. 
    
    \item A color-color diagram was created to determine the masses and the ages of clusters using the SSP models. To see the age-extinction degeneracy due to the contribution of nebular emission, we also included data from F658N filter. We saw that, some clusters in the high extinction regions tend to be younger, while some younger clusters in low nebular emission to be older.

    \item We identified a total of 42 XRBs within the {\it HST}/ACS fields. 15 of these (36\%) were classified as HMXBs based on having optical counterpart(s). The remaining 27 sources (64\%) were classified as LMXBs due to their lack of optical counterparts in the {\it HST} images.
    
    \item The XLFs of XRBs were then obtained and fitted with a power-law model. The slopes for HMXBs and LMXBs were determined as $\alpha$=-1.50$\pm0.07$ and $\alpha$=-1.70$\pm0.15$, respectively.
    
    \item 18 of XRBs are associated with star clusters within their 200 pc. While 5 of these sources are located close to very young clusters ($\tau$ < 10$^{7}$ yr), and 13 are located close to older clusters aged $\tau$ > 10$^{8}$ yr.
    
    \item The spatial distributions of the XRBs show that the HMXBs are located in the spiral arms as expected. However, some LMXBs are located in the galaxy's bulge, but with no significant concentration and are generally distributed throughout the galactic plane.
    
    \item The host clusters of HMXBs are young, while those of LMXBs are older. However, three of the sources classified as LMXB, with associated clusters are younger than 10 Myr. This ambiguity may be due to a chance alignment for the source (X37) or  a difficulty in detecting the optical counterparts of two sources (X46 and X55).

    
    \item The age-mass distribution of the  identified star clusters indicates that massive star clusters can evolve to older ages than low-mass clusters. This behavior is in line with what is expected of star clusters in a galaxy. In addition, this distribution in NGC 628 is well explained by the mass-independent disruption model.
    
   \item Among the 42 XRBs, 15 LMXBs within the {\it HST}/ACS fields do not associate with any star cluster in the region < 200 pc radius. 
   Since these host clusters seem to have low-masses 
  they might have disrupted when they were young. 
  Also, 9 out of 15 identified donor stars (classified as HMXBs) were not associated with any star clusters. If these 9 sources were also born in  clusters, they are expected to be  young (<10$^{8.5}$yr). This means, considering the mass-age distribution at hand, the young clusters below an approximate mass (< 10$^{3.7}$$M_{\odot}$) must have dispersed before evolving into older ages. 
    

\end{enumerate}
We may tentatively conclude that about 1/3 of XRBs are HMXBs, while 2/3 are LMXBs, and cluster association statistics for these sources give a ratio of about 1/2 for the association/no association division.
All these rough ratios could provide us hints for the formation and evolutionary history of XRBs and nearby clusters. This may need further elaboration of multiwavelength studies of such populations in nearby galaxies.

\section{ACKNOWLEDGEMENTS}
This research was supported by the Scientific and Technological Research Council of Turkey (TÜBİTAK) through project number 119F315. We thank Prof. M. E. Özel for his valuable contributions and suggestions. We acknowledge the valuable comments and recommendations of the Reviewer which helped to clarify a number of issues.

\section*{Data Availability}
The scientific results reported in this article are based on archival observations made by the {\it Chandra}\footnote{https://cda.harvard.edu/chaser/} X-ray Observatories. This work has also made use of observations made with the NASA/ESA {\it Hubble Space Telescope}, and obtained from the data archive at the Space Telescope Science Institute\footnote{https://mast.stsci.edu/search/ui/$\#$/hst}

\bibliographystyle{mnras}
\bibliography{Avdan_et_al} 

\begin{thebibliography}{}
\makeatletter
\relax
\def\mn@urlcharsother{\let\do\@makeother \do\$\do\&\do\#\do\^\do\_\do\%\do\~}
\def\mn@doi{\begingroup\mn@urlcharsother \@ifnextchar [ {\mn@doi@}
  {\mn@doi@[]}}
\def\mn@doi@[#1]#2{\def\@tempa{#1}\ifx\@tempa\@empty \href
  {http://dx.doi.org/#2} {doi:#2}\else \href {http://dx.doi.org/#2} {#1}\fi
  \endgroup}
\def\mn@eprint#1#2{\mn@eprint@#1:#2::\@nil}
\def\mn@eprint@arXiv#1{\href {http://arxiv.org/abs/#1} {{\tt arXiv:#1}}}
\def\mn@eprint@dblp#1{\href {http://dblp.uni-trier.de/rec/bibtex/#1.xml}
  {dblp:#1}}
\def\mn@eprint@#1:#2:#3:#4\@nil{\def\@tempa {#1}\def\@tempb {#2}\def\@tempc
  {#3}\ifx \@tempc \@empty \let \@tempc \@tempb \let \@tempb \@tempa \fi \ifx
  \@tempb \@empty \def\@tempb {arXiv}\fi \@ifundefined
  {mn@eprint@\@tempb}{\@tempb:\@tempc}{\expandafter \expandafter \csname
  mn@eprint@\@tempb\endcsname \expandafter{\@tempc}}}

\bibitem[\protect\citeauthoryear{{Adamo} et~al.,}{{Adamo}
  et~al.}{2017}]{2017ApJ...841..131A}
{Adamo} A.,  et~al., 2017, \mn@doi [\apj] {10.3847/1538-4357/aa7132}, \href
  {https://ui.adsabs.harvard.edu/abs/2017ApJ...841..131A} {841, 131}

\bibitem[\protect\citeauthoryear{{Allak}}{{Allak}}{2022}]{2022MNRAS.517.3495A}
{Allak} S.,  2022, \mn@doi [\mnras] {10.1093/mnras/stac2887}, \href
  {https://ui.adsabs.harvard.edu/abs/2022MNRAS.517.3495A} {517, 3495}

\bibitem[\protect\citeauthoryear{{Allak}, {Akyuz}, {Sonbas}  \&
  {Dhuga}}{{Allak} et~al.}{2022}]{2022MNRAS.515.3632A}
{Allak} S.,  {Akyuz} A.,  {Sonbas} E.,   {Dhuga} K.~S.,  2022, \mn@doi [\mnras]
  {10.1093/mnras/stac1992}, \href
  {https://ui.adsabs.harvard.edu/abs/2022MNRAS.515.3632A} {515, 3632}

\bibitem[\protect\citeauthoryear{{Aller} et~al.,}{{Aller}
  et~al.}{1982}]{1982lbg6.conf.....A}
{Aller} L.~H.,  et~al., eds, 1982, {Landolt-B{\"o}rnstein: Numerical Data and
  Functional Relationships in Science and Technology - New Series ``
  Gruppe/Group 6 Astronomy and Astrophysics '' Volume 2 Schaifers/Voigt:
  Astronomy and Astrophysics / Astronomie und Astrophysik `` Stars and Star
  Clusters / Sterne und Sternhaufen}

\bibitem[\protect\citeauthoryear{{Bahramian} \& {Degenaar}}{{Bahramian} \&
  {Degenaar}}{2022}]{2022arXiv220610053B}
{Bahramian} A.,  {Degenaar} N.,  2022, arXiv e-prints, \href
  {https://ui.adsabs.harvard.edu/abs/2022arXiv220610053B} {p. arXiv:2206.10053}

\bibitem[\protect\citeauthoryear{{Bastian} et~al.,}{{Bastian}
  et~al.}{2012}]{2012MNRAS.419.2606B}
{Bastian} N.,  et~al., 2012, \mn@doi [\mnras]
  {10.1111/j.1365-2966.2011.19909.x}, \href
  {https://ui.adsabs.harvard.edu/abs/2012MNRAS.419.2606B} {419, 2606}

\bibitem[\protect\citeauthoryear{{Bruzual} \& {Charlot}}{{Bruzual} \&
  {Charlot}}{2003}]{2003MNRAS.344..1000B}
{Bruzual} G.,  {Charlot} S.,  2003, \mn@doi [\mnras]
  {10.1046/j.1365-8711.2003.06897.x}, \href
  {http://www.iap.fr/~charlot/bc2003/paper} {344, 1000B}

\bibitem[\protect\citeauthoryear{{Chandar}, {Fall}  \& {Whitmore}}{{Chandar}
  et~al.}{2010a}]{2010ApJ...711.1263C}
{Chandar} R.,  {Fall} S.~M.,   {Whitmore} B.~C.,  2010a, \mn@doi [\apj]
  {10.1088/0004-637X/711/2/1263}, \href
  {https://ui.adsabs.harvard.edu/abs/2010ApJ...711.1263C} {711, 1263}

\bibitem[\protect\citeauthoryear{{Chandar}, {Whitmore}, {Kim}  \& et
  al.}{{Chandar} et~al.}{2010b}]{2010ApJ...719..966C}
{Chandar} R.,  {Whitmore} B.~C.,  {Kim} H.,   et al. 2010b, \mn@doi [\apj]
  {10.1088/0004-637X/719/1/966}, \href
  {https://ui.adsabs.harvard.edu/abs/2010ApJ...719..966C/abstract} {719, 966}

\bibitem[\protect\citeauthoryear{{Chandar}, {Whitmore}, {Dinino}, {Kennicutt},
  {Chien}, {Schinnerer}  \& {Meidt}}{{Chandar}
  et~al.}{2016}]{2016ApJ...824...71C}
{Chandar} R.,  {Whitmore} B.~C.,  {Dinino} D.,  {Kennicutt} R.~C.,  {Chien}
  L.~H.,  {Schinnerer} E.,   {Meidt} S.,  2016, \mn@doi [\apj]
  {10.3847/0004-637X/824/2/71}, \href
  {https://ui.adsabs.harvard.edu/abs/2016ApJ...824...71C} {824, 71}

\bibitem[\protect\citeauthoryear{{Chandar}, {Johns}, {Mok}, {Prestwich},
  {Gallo}  \& {Hunt}}{{Chandar} et~al.}{2020}]{2020ApJ...890..150C}
{Chandar} R.,  {Johns} P.,  {Mok} A.,  {Prestwich} A.,  {Gallo} E.,   {Hunt}
  Q.,  2020, \mn@doi [\apj] {10.3847/1538-4357/ab6b27}, \href
  {https://ui.adsabs.harvard.edu/abs/2020ApJ...890..150C} {890, 150}

\bibitem[\protect\citeauthoryear{{Dickey} \& {Lockman}}{{Dickey} \&
  {Lockman}}{1990}]{1990ARA&A..28..215D}
{Dickey} J.~M.,  {Lockman} F.~J.,  1990, \mn@doi [\araa]
  {10.1146/annurev.aa.28.090190.001243}, \href
  {https://ui.adsabs.harvard.edu/abs/1990ARA&A..28..215D} {28, 215}

\bibitem[\protect\citeauthoryear{{Elson}, {Fall}  \& {Freeman}}{{Elson}
  et~al.}{1987}]{1987ApJ...323...54E}
{Elson} R. A.~W.,  {Fall} S.~M.,   {Freeman} K.~C.,  1987, \mn@doi [\apj]
  {10.1086/165807}, \href
  {https://ui.adsabs.harvard.edu/abs/1987ApJ...323...54E/abstract} {323, 54}

\bibitem[\protect\citeauthoryear{{Ertan} \& {Alpar}}{{Ertan} \&
  {Alpar}}{2002}]{2002A&A...393..205E}
{Ertan} {\"U}.,  {Alpar} M.~A.,  2002, \mn@doi [\aap]
  {10.1051/0004-6361:20020998}, \href
  {https://ui.adsabs.harvard.edu/abs/2002A&A...393..205E} {393, 205}

\bibitem[\protect\citeauthoryear{{Fabbiano}}{{Fabbiano}}{1989}]{1989ARA&A..27...87F}
{Fabbiano} G.,  1989, \mn@doi [\araa] {10.1146/annurev.aa.27.090189.000511},
  \href {https://ui.adsabs.harvard.edu/abs/1989ARA&A..27...87F} {27, 87}

\bibitem[\protect\citeauthoryear{{Fabbiano}}{{Fabbiano}}{2006}]{2006ARA&A..44..323F}
{Fabbiano} G.,  2006, \mn@doi [\araa] {10.1146/annurev.astro.44.051905.092519},
  \href {https://ui.adsabs.harvard.edu/abs/2006ARA&A..44..323F} {44, 323}

\bibitem[\protect\citeauthoryear{{Fall}, {Chandar}  \& {Whitmore}}{{Fall}
  et~al.}{2009}]{2009ApJ...704..453F}
{Fall} S.~M.,  {Chandar} R.,   {Whitmore} B.~C.,  2009, \mn@doi [\apj]
  {10.1088/0004-637X/704/1/453}, \href
  {https://ui.adsabs.harvard.edu/abs/2009ApJ...704..453F} {704, 453}

\bibitem[\protect\citeauthoryear{{Ferrigno}, {Farinelli}, {Bozzo},
  {Pottschmidt}, {Klochkov}  \& {Kretschmar}}{{Ferrigno}
  et~al.}{2013}]{2013A&A...553A.103F}
{Ferrigno} C.,  {Farinelli} R.,  {Bozzo} E.,  {Pottschmidt} K.,  {Klochkov} D.,
    {Kretschmar} P.,  2013, \mn@doi [\aap] {10.1051/0004-6361/201321053}, \href
  {https://ui.adsabs.harvard.edu/abs/2013A&A...553A.103F} {553, A103}

\bibitem[\protect\citeauthoryear{{Flesch}}{{Flesch}}{2021}]{2021yCat.7290....0F}
{Flesch} E.~W.,  2021, VizieR Online Data Catalog, \href
  {https://ui.adsabs.harvard.edu/abs/2021yCat.7290....0F} {p. VII/290}

\bibitem[\protect\citeauthoryear{{Garofali}, {Converse}, {Chandar}  \&
  {Rangelov}}{{Garofali} et~al.}{2012}]{2012ApJ...755...49G}
{Garofali} K.,  {Converse} J.~M.,  {Chandar} R.,   {Rangelov} B.,  2012,
  \mn@doi [\apj] {10.1088/0004-637X/755/1/49}, \href
  {https://ui.adsabs.harvard.edu/abs/2012ApJ...755...49G} {755, 49}

\bibitem[\protect\citeauthoryear{{Gilfanov}}{{Gilfanov}}{2004}]{2004MNRAS.349..146G}
{Gilfanov} M.,  2004, \mn@doi [\mnras] {10.1111/j.1365-2966.2004.07473.x},
  \href {https://ui.adsabs.harvard.edu/abs/2004MNRAS.349..146G} {349, 146}

\bibitem[\protect\citeauthoryear{{Gilfanov}, {Grimm}  \& {Sunyaev}}{{Gilfanov}
  et~al.}{2004}]{2004MNRAS.347L..57G}
{Gilfanov} M.,  {Grimm} H.~J.,   {Sunyaev} R.,  2004, \mn@doi [\mnras]
  {10.1111/j.1365-2966.2004.07450.x}, \href
  {https://ui.adsabs.harvard.edu/abs/2004MNRAS.347L..57G} {347, L57}

\bibitem[\protect\citeauthoryear{{Grasha} et~al.,}{{Grasha}
  et~al.}{2015}]{2015ApJ...815...93G}
{Grasha} K.,  et~al., 2015, \mn@doi [\apj] {10.1088/0004-637X/815/2/93}, \href
  {https://ui.adsabs.harvard.edu/abs/2015ApJ...815...93G} {815, 93}

\bibitem[\protect\citeauthoryear{{Grimm}, {Gilfanov}  \& {Sunyaev}}{{Grimm}
  et~al.}{2003a}]{2003ChJAS...3..257G}
{Grimm} H.-J.,  {Gilfanov} M.,   {Sunyaev} R.,  2003a, \mn@doi [Chinese Journal
  of Astronomy and Astrophysics Supplement] {10.1088/1009-9271/3/S1/257}, \href
  {https://ui.adsabs.harvard.edu/abs/2003ChJAS...3..257G} {3, 257}

\bibitem[\protect\citeauthoryear{{Grimm}, {Gilfanov}  \& {Sunyaev}}{{Grimm}
  et~al.}{2003b}]{2003MNRAS.339..793G}
{Grimm} H.~J.,  {Gilfanov} M.,   {Sunyaev} R.,  2003b, \mn@doi [\mnras]
  {10.1046/j.1365-8711.2003.06224.x}, \href
  {https://ui.adsabs.harvard.edu/abs/2003MNRAS.339..793G} {339, 793}

\bibitem[\protect\citeauthoryear{{Haberl}, {Maitra}, {Vasilopoulos}, {Maggi},
  {Udalski}, {Monageng}  \& {Buckley}}{{Haberl}
  et~al.}{2022}]{2022A&A...662A..22H}
{Haberl} F.,  {Maitra} C.,  {Vasilopoulos} G.,  {Maggi} P.,  {Udalski} A.,
  {Monageng} I.~M.,   {Buckley} D.~A.~H.,  2022, \mn@doi [\aap]
  {10.1051/0004-6361/202243301}, \href
  {https://ui.adsabs.harvard.edu/abs/2022A&A...662A..22H} {662, A22}

\bibitem[\protect\citeauthoryear{{Hameury}}{{Hameury}}{2020}]{2020AdSpR..66.1004H}
{Hameury} J.~M.,  2020, \mn@doi [Advances in Space Research]
  {10.1016/j.asr.2019.10.022}, \href
  {https://ui.adsabs.harvard.edu/abs/2020AdSpR..66.1004H} {66, 1004}

\bibitem[\protect\citeauthoryear{{Hofmann}, {Pietsch}, {Henze}, {Haberl},
  {Sturm}, {Della Valle}, {Hartmann}  \& {Hatzidimitriou}}{{Hofmann}
  et~al.}{2013}]{2013A&A...555A..65H}
{Hofmann} F.,  {Pietsch} W.,  {Henze} M.,  {Haberl} F.,  {Sturm} R.,  {Della
  Valle} M.,  {Hartmann} D.~H.,   {Hatzidimitriou} D.,  2013, \mn@doi [\aap]
  {10.1051/0004-6361/201321165}, \href
  {https://ui.adsabs.harvard.edu/abs/2013A&A...555A..65H} {555, A65}

\bibitem[\protect\citeauthoryear{{Hunt}, {Gallo}, {Chandar}, {Johns Mulia},
  {Mok}, {Prestwich}  \& {Liu}}{{Hunt} et~al.}{2021}]{2021ApJ...912...31H}
{Hunt} Q.,  {Gallo} E.,  {Chandar} R.,  {Johns Mulia} P.,  {Mok} A.,
  {Prestwich} A.,   {Liu} S.,  2021, \mn@doi [\apj] {10.3847/1538-4357/abe531},
  \href {https://ui.adsabs.harvard.edu/abs/2021ApJ...912...31H} {912, 31}

\bibitem[\protect\citeauthoryear{{Irwin}}{{Irwin}}{2006}]{2006MNRAS.371.1903I}
{Irwin} J.~A.,  2006, \mn@doi [\mnras] {10.1111/j.1365-2966.2006.10822.x},
  \href {https://ui.adsabs.harvard.edu/abs/2006MNRAS.371.1903I} {371, 1903}

\bibitem[\protect\citeauthoryear{{Jin} \& {Kong}}{{Jin} \&
  {Kong}}{2019}]{2019ApJ...879..112J}
{Jin} R.,  {Kong} A. K.~H.,  2019, \mn@doi [\apj] {10.3847/1538-4357/ab2461},
  \href {https://ui.adsabs.harvard.edu/abs/2019ApJ...879..112J} {879, 112}

\bibitem[\protect\citeauthoryear{{Johns Mulia}, {Chandar}  \&
  {Rangelov}}{{Johns Mulia} et~al.}{2019}]{2019ApJ...871..122J}
{Johns Mulia} P.,  {Chandar} R.,   {Rangelov} B.,  2019, \mn@doi [\apj]
  {10.3847/1538-4357/aaf56a}, \href
  {https://ui.adsabs.harvard.edu/abs/2019ApJ...871..122J} {871, 122}

\bibitem[\protect\citeauthoryear{{Johnson} et~al.,}{{Johnson}
  et~al.}{2012}]{2012ApJ...752...95J}
{Johnson} C.~L.,  et~al., 2012, \mn@doi [\aap] {10.1088/0004-637X/752/2/95},
  \href {https://ui.adsabs.harvard.edu/abs/2012ApJ...752...95J/abstract} {752,
  23}

\bibitem[\protect\citeauthoryear{{Kaaret}, {Alonso-Herrero}, {Gallagher},
  {Fabbiano}, {Zezas}  \& {Rieke}}{{Kaaret} et~al.}{2004}]{2004MNRAS.348L..28K}
{Kaaret} P.,  {Alonso-Herrero} A.,  {Gallagher} J.~S.,  {Fabbiano} G.,  {Zezas}
  A.,   {Rieke} M.~J.,  2004, \mn@doi [\mnras]
  {10.1111/j.1365-2966.2004.07516.x}, \href
  {https://ui.adsabs.harvard.edu/abs/2004MNRAS.348L..28K} {348, L28}

\bibitem[\protect\citeauthoryear{{Kaaret}, {Feng}  \& {Roberts}}{{Kaaret}
  et~al.}{2017}]{2017ARA&A..55..303K}
{Kaaret} P.,  {Feng} H.,   {Roberts} T.~P.,  2017, \mn@doi [\araa]
  {10.1146/annurev-astro-091916-055259}, \href
  {https://ui.adsabs.harvard.edu/abs/2017ARA&A..55..303K} {55, 303}

\bibitem[\protect\citeauthoryear{{Krauss}, {Kilgard}, {Garcia}, {Roberts}  \&
  {Prestwich}}{{Krauss} et~al.}{2005}]{2005ApJ...630..228K}
{Krauss} M.~I.,  {Kilgard} R.~E.,  {Garcia} M.~R.,  {Roberts} T.~P.,
  {Prestwich} A.~H.,  2005, \mn@doi [\apj] {10.1086/431784}, \href
  {https://ui.adsabs.harvard.edu/abs/2005ApJ...630..228K} {630, 228}

\bibitem[\protect\citeauthoryear{{Larsen}}{{Larsen}}{1999}]{1999A&AS..139..393L}
{Larsen} S.~S.,  1999, \mn@doi [\aaps] {10.1051/aas:1999509}, \href
  {https://ui.adsabs.harvard.edu/abs/1999A&AS..139..393L} {139, 393}

\bibitem[\protect\citeauthoryear{{Larsen}}{{Larsen}}{2002}]{2002AJ....124.1393L}
{Larsen} S.~S.,  2002, \mn@doi [\aj] {10.1086/342381}, \href
  {https://ui.adsabs.harvard.edu/abs/2002AJ....124.1393L} {124, 1393}

\bibitem[\protect\citeauthoryear{{Larsen}}{{Larsen}}{2004}]{2004A&A...416..537L}
{Larsen} S.~S.,  2004, \mn@doi [\aap] {10.1051/0004-6361:20034533}, \href
  {https://ui.adsabs.harvard.edu/abs/2004A&A...416..537L} {416, 537}

\bibitem[\protect\citeauthoryear{{Larsen} \& {Brodie}}{{Larsen} \&
  {Brodie}}{2000}]{2000Aj..120..2938}
{Larsen} S.,  {Brodie} J.,  2000, \mn@doi [\aj] {10.1086/316847}, \href
  {https://ui.adsabs.harvard.edu/abs/2000AJ....120.2938L/abstract} {120, 2938}

\bibitem[\protect\citeauthoryear{{Larsen} \& {Richtler}}{{Larsen} \&
  {Richtler}}{1999}]{1999A&A...345...59L}
{Larsen} S.~S.,  {Richtler} T.,  1999, \aap, \href
  {https://ui.adsabs.harvard.edu/abs/1999A&A...345...59L} {345, 59}

\bibitem[\protect\citeauthoryear{{Lehmer} et~al.,}{{Lehmer}
  et~al.}{2019}]{2019ApJS..243....3L}
{Lehmer} B.~D.,  et~al., 2019, \mn@doi [\apjs] {10.3847/1538-4365/ab22a8},
  \href {https://ui.adsabs.harvard.edu/abs/2019ApJS..243....3L} {243, 3}

\bibitem[\protect\citeauthoryear{{Lim}, {Hwang}  \& {Lee}}{{Lim}
  et~al.}{2013}]{2013ApJ...766...20L}
{Lim} S.,  {Hwang} N.,   {Lee} M.~G.,  2013, \mn@doi [\apj]
  {10.1088/0004-637X/766/1/20}, \href
  {https://ui.adsabs.harvard.edu/abs/2013ApJ...766...20L} {766, 20}

\bibitem[\protect\citeauthoryear{{Lomel{\'\i}-N{\'u}{\~n}ez}, {Mayya},
  {Rodr{\'\i}guez-Merino}, {Ovando}  \&
  {Rosa-Gonz{\'a}lez}}{{Lomel{\'\i}-N{\'u}{\~n}ez}
  et~al.}{2022}]{2022MNRAS.509..180L}
{Lomel{\'\i}-N{\'u}{\~n}ez} L.,  {Mayya} Y.~D.,  {Rodr{\'\i}guez-Merino} L.~H.,
   {Ovando} P.~A.,   {Rosa-Gonz{\'a}lez} D.,  2022, \mn@doi [\mnras]
  {10.1093/mnras/stab2890}, \href
  {https://ui.adsabs.harvard.edu/abs/2022MNRAS.509..180L} {509, 180}

\bibitem[\protect\citeauthoryear{{Mineo}, {Gilfanov}  \& {Sunyaev}}{{Mineo}
  et~al.}{2012}]{2012MNRAS.419.2095M}
{Mineo} S.,  {Gilfanov} M.,   {Sunyaev} R.,  2012, \mn@doi [\mnras]
  {10.1111/j.1365-2966.2011.19862.x}, \href
  {https://ui.adsabs.harvard.edu/abs/2012MNRAS.419.2095M} {419, 2095}

\bibitem[\protect\citeauthoryear{{Mineo}, {Gilfanov}, {Lehmer}, {Morrison}  \&
  {Sunyaev}}{{Mineo} et~al.}{2014}]{2014MNRAS.437.1698M}
{Mineo} S.,  {Gilfanov} M.,  {Lehmer} B.~D.,  {Morrison} G.~E.,   {Sunyaev} R.,
   2014, \mn@doi [\mnras] {10.1093/mnras/stt1999}, \href
  {https://ui.adsabs.harvard.edu/abs/2014MNRAS.437.1698M} {437, 1698}

\bibitem[\protect\citeauthoryear{{Mora}, {Larsen}, {Kissler-Patig}, {Brodie}
  \& {Richtler}}{{Mora} et~al.}{2009}]{2009A&A...501..949M}
{Mora} M.~D.,  {Larsen} S.~S.,  {Kissler-Patig} M.,  {Brodie} J.~P.,
  {Richtler} T.,  2009, \mn@doi [\aap] {10.1051/0004-6361/200810614}, \href
  {https://ui.adsabs.harvard.edu/abs/2009A&A...501..949M} {501, 949}

\bibitem[\protect\citeauthoryear{{Motch} et~al.,}{{Motch}
  et~al.}{2016}]{2016yCat.9048....0M}
{Motch} C.,  et~al., 2016, VizieR Online Data Catalog, \href
  {https://ui.adsabs.harvard.edu/abs/2016yCat.9048....0M} {p. IX/48}

\bibitem[\protect\citeauthoryear{{Mulcahy}, {Beck}  \& {Heald}}{{Mulcahy}
  et~al.}{2017}]{2017A&A...600A...6M}
{Mulcahy} D.~D.,  {Beck} R.,   {Heald} G.~H.,  2017, \mn@doi [\aap]
  {10.1051/0004-6361/201629907}, \href
  {https://ui.adsabs.harvard.edu/abs/2017A&A...600A...6M} {600, A6}

\bibitem[\protect\citeauthoryear{{Prasow-{\'E}mond} et~al.,}{{Prasow-{\'E}mond}
  et~al.}{2022}]{2022AJ....164....7P}
{Prasow-{\'E}mond} M.,  et~al., 2022, \mn@doi [\aj] {10.3847/1538-3881/ac6d57},
  \href {https://ui.adsabs.harvard.edu/abs/2022AJ....164....7P} {164, 7}

\bibitem[\protect\citeauthoryear{{Prestwich}, {Irwin}, {Kilgard}, {Krauss},
  {Zezas}, {Primini}, {Kaaret}  \& {Boroson}}{{Prestwich}
  et~al.}{2003}]{2003ApJ...595..719P}
{Prestwich} A.~H.,  {Irwin} J.~A.,  {Kilgard} R.~E.,  {Krauss} M.~I.,  {Zezas}
  A.,  {Primini} F.,  {Kaaret} P.,   {Boroson} B.,  2003, \mn@doi [\apj]
  {10.1086/377366}, \href
  {https://ui.adsabs.harvard.edu/abs/2003ApJ...595..719P} {595, 719}

\bibitem[\protect\citeauthoryear{{Rangelov}, {Prestwich}  \&
  {Chandar}}{{Rangelov} et~al.}{2011}]{2011ApJ...741...86R}
{Rangelov} B.,  {Prestwich} A.~H.,   {Chandar} R.,  2011, \mn@doi [\apj]
  {10.1088/0004-637X/741/2/86}, \href
  {https://ui.adsabs.harvard.edu/abs/2011ApJ...741...86R} {741, 86}

\bibitem[\protect\citeauthoryear{{Reig}}{{Reig}}{2011}]{2011Ap&SS.332....1R}
{Reig} P.,  2011, \mn@doi [\apss] {10.1007/s10509-010-0575-8}, \href
  {https://ui.adsabs.harvard.edu/abs/2011Ap&SS.332....1R} {332, 1}

\bibitem[\protect\citeauthoryear{{Remillard} \& {McClintock}}{{Remillard} \&
  {McClintock}}{2006}]{2006ARA&A..44...49R}
{Remillard} R.~A.,  {McClintock} J.~E.,  2006, \mn@doi [\araa]
  {10.1146/annurev.astro.44.051905.092532}, \href
  {https://ui.adsabs.harvard.edu/abs/2006ARA&A..44...49R} {44, 49}

\bibitem[\protect\citeauthoryear{{Rice}, {Rangelov}, {Prestwich}, {Chandar},
  {Bichon}  \& {Boldt}}{{Rice} et~al.}{2021}]{2021ApJ...922..178R}
{Rice} J.~R.,  {Rangelov} B.,  {Prestwich} A.,  {Chandar} R.,  {Bichon} L.,
  {Boldt} C.,  2021, \mn@doi [\apj] {10.3847/1538-4357/ac22ac}, \href
  {https://ui.adsabs.harvard.edu/abs/2021ApJ...922..178R} {922, 178}

\bibitem[\protect\citeauthoryear{{Russell}, {Fender}, {Hynes}, {Brocksopp},
  {Homan}, {Jonker}  \& {Buxton}}{{Russell} et~al.}{2006}]{2006MNRAS.371.1334R}
{Russell} D.~M.,  {Fender} R.~P.,  {Hynes} R.~I.,  {Brocksopp} C.,  {Homan} J.,
   {Jonker} P.~G.,   {Buxton} M.~M.,  2006, \mn@doi [\mnras]
  {10.1111/j.1365-2966.2006.10756.x}, \href
  {https://ui.adsabs.harvard.edu/abs/2006MNRAS.371.1334R} {371, 1334}

\bibitem[\protect\citeauthoryear{{Ryon } et~al.,}{{Ryon }
  et~al.}{2017}]{2017ApJ...841...92R}
{Ryon } J.~E.,  et~al., 2017, \mn@doi [\aap] {10.3847/1538-4357/aa719e}, \href
  {https://ui.adsabs.harvard.edu/abs/2017ApJ...841...92R/abstract} {841, Issue
  2, article id. 92, 16 pp.}

\bibitem[\protect\citeauthoryear{{Sasaki} et~al.,}{{Sasaki}
  et~al.}{2018}]{2018A&A...620A..28S}
{Sasaki} M.,  et~al., 2018, \mn@doi [\aap] {10.1051/0004-6361/201833588}, \href
  {https://ui.adsabs.harvard.edu/abs/2018A&A...620A..28S} {620, A28}

\bibitem[\protect\citeauthoryear{{Sirianni} et~al.,}{{Sirianni}
  et~al.}{2005}]{2005PASP..117.1049S}
{Sirianni} M.,  et~al., 2005, \mn@doi [ASP] {10.1086/444553}, \href
  {https://ui.adsabs.harvard.edu/abs/2005PASP..117.1049S/abstract} {117, 1049}

\bibitem[\protect\citeauthoryear{{Sonba{\c{s}}}, {Aky{\"u}z}, {Balman}  \&
  {{\"O}zel}}{{Sonba{\c{s}}} et~al.}{2010}]{2010A&A...517A..91S}
{Sonba{\c{s}}} E.,  {Aky{\"u}z} A.,  {Balman} {\c{S}}.,   {{\"O}zel} M.~E.,
  2010, \mn@doi [\aap] {10.1051/0004-6361/200913858}, \href
  {https://ui.adsabs.harvard.edu/abs/2010A&A...517A..91S} {517, A91}

\bibitem[\protect\citeauthoryear{{Tauris} \& {van den Heuvel}}{{Tauris} \& {van
  den Heuvel}}{2006}]{2006csxs.book..623T}
{Tauris} T.~M.,  {van den Heuvel} E.~P.~J.,  2006, in , Vol.~39, Compact
  stellar X-ray sources.
Cambridge University Press, pp 623--665

\bibitem[\protect\citeauthoryear{{Tully}}{{Tully}}{1988}]{1988Sci...242..310T}
{Tully} R.~B.,  1988, Science, \href
  {https://ui.adsabs.harvard.edu/abs/1988Sci...242..310T} {242, 310}

\bibitem[\protect\citeauthoryear{{Ujjwal}, {Kartha}, {Subramanian}, {George},
  {Thomas}  \& {Mathew}}{{Ujjwal} et~al.}{2022}]{2022MNRAS.516.2171U}
{Ujjwal} K.,  {Kartha} S.~S.,  {Subramanian} S.,  {George} K.,  {Thomas} R.,
  {Mathew} B.,  2022, \mn@doi [\mnras] {10.1093/mnras/stac2285}, \href
  {https://ui.adsabs.harvard.edu/abs/2022MNRAS.516.2171U} {516, 2171}

\bibitem[\protect\citeauthoryear{{Vulic}, {Barmby}  \& {Gallagher}}{{Vulic}
  et~al.}{2013}]{2013ApJ...763...96V}
{Vulic} N.,  {Barmby} P.,   {Gallagher} S.~C.,  2013, \mn@doi [\apj]
  {10.1088/0004-637X/763/2/96}, \href
  {https://ui.adsabs.harvard.edu/abs/2013ApJ...763...96V} {763, 96}

\bibitem[\protect\citeauthoryear{{Vulic} et~al.,}{{Vulic}
  et~al.}{2018}]{2018ApJ...864..150V}
{Vulic} N.,  et~al., 2018, \mn@doi [\apj] {10.3847/1538-4357/aad500}, \href
  {https://ui.adsabs.harvard.edu/abs/2018ApJ...864..150V} {864, 150}

\bibitem[\protect\citeauthoryear{{Whitmore}, {Chandar}  \& {Fall}}{{Whitmore}
  et~al.}{2007}]{2007AJ....133.1067W}
{Whitmore} B.~C.,  {Chandar} R.,   {Fall} S.~M.,  2007, \mn@doi [\aj]
  {10.1086/510288}, \href
  {https://ui.adsabs.harvard.edu/abs/2007AJ....133.1067W} {133, 1067}

\bibitem[\protect\citeauthoryear{{Whitmore}, {Chandar}, {Bowers}, {Larsen},
  {Lindsay}, {Ansari}  \& {Evans}}{{Whitmore}
  et~al.}{2014}]{2014AJ....147...78W}
{Whitmore} B.~C.,  {Chandar} R.,  {Bowers} A.~S.,  {Larsen} S.,  {Lindsay} K.,
  {Ansari} A.,   {Evans} J.,  2014, \mn@doi [\aj] {10.1088/0004-6256/147/4/78},
  \href {https://ui.adsabs.harvard.edu/abs/2014AJ....147...78W} {147, 78}

\bibitem[\protect\citeauthoryear{{Zezas}, {Fabbiano}, {Rots}  \&
  {Murray}}{{Zezas} et~al.}{2002}]{2002ApJ...577..710Z}
{Zezas} A.,  {Fabbiano} G.,  {Rots} A.~H.,   {Murray} S.~S.,  2002, \mn@doi
  [\apj] {10.1086/342160}, \href
  {https://ui.adsabs.harvard.edu/abs/2002ApJ...577..710Z} {577, 710}

\bibitem[\protect\citeauthoryear{{Zhang}, {Gilfanov}  \& {Bogd{\'a}n}}{{Zhang}
  et~al.}{2013}]{2013A&A...556A...9Z}
{Zhang} Z.,  {Gilfanov} M.,   {Bogd{\'a}n} {\'A}.,  2013, \mn@doi [\aap]
  {10.1051/0004-6361/201220685}, \href
  {https://ui.adsabs.harvard.edu/abs/2013A&A...556A...9Z} {556, A9}

\bibitem[\protect\citeauthoryear{{Zou} et~al.,}{{Zou}
  et~al.}{2011}]{2011AJ....142...16Z}
{Zou} H.,  et~al., 2011, \mn@doi [\aj] {10.1088/0004-6256/142/1/16}, \href
  {https://ui.adsabs.harvard.edu/abs/2011AJ....142...16Z} {142, 16}

\makeatother
\end{thebibliography}

\begin{table*}
\centering
\caption{The log of {\it Chandra} observations of NGC 628}
\begin{tabular}{lccrcc}
\hline\hline
Label & ObsID & Date & Exp.  \\
 & & & (ks)  \\
\hline
Ch1 & 2057 & 2001-06-19 &  46.36  \\
Ch2 & 2058& 2001-10-19 & 46.17  \\
Ch3 & 4753& 2003-11-20 & 5.28  \\
Ch4 & 4754& 2003-12-29 & 5.04  \\
Ch5 & 14801& 2013-08-21 & 9.84 \\
Ch6 & 16000 & 2013-09-21 & 39.55 \\
Ch7 & 16001& 2013-10-07 & 14.73 \\
Ch8 & 16484& 2013-10-10 & 14.68 \\
Ch9 & 16485& 2013-10-11 & 8.99 \\
Ch10 & 16002 & 2013-11-14 & 37.57 \\
Ch11 & 16003 & 2013-12-15 & 40.44 \\
Ch12 & 21000& 2018-09-30 & 9.96 \\
Ch13 & 20333 & 2018-11-04 & 15.84\\
\hline
\end{tabular}
\label{T:x-ray}
\end{table*}

\begin{table*}
\centering
\caption{The best-fit spectral model parameters of X-ray sources with >50 counts in NGC 628. X-ray luminosities were calculated in 0.3-8 keV energy range.}
\begin{tabular}{c c c c c r c c c c }
\hline
ID   & RA & Dec. & model& $\Gamma$ & kT$_{\mathrm{in}}$ & $\chi^{2}$/dof &  L$_{\mathrm{X}}$         \\
     & (deg.) & (deg.)    &    &  &  (keV) &   &   10$^{38}$ (erg s$^{-1}$)   \\
\hline

X1	& 24.212833 & 15.763083 & po+diskbb & $1.12_{-0.09}^{+0.10}$ & $0.28_{-0.01}^{+0.01}$ & 49.57/49 & $18.72_{-0.03}^{+0.03}$ \\
X2 & 24.183875 & 15.805208	& diskbb & $-$ &  $1.27_{-0.04}^{+0.04}$ &  19.96/27 & $5.12_{-0.05}^{+0.04}$  \\
X3 & 24.163083 & 15.719466	& po & $1.59_{-0.16}^{+0.19}$ & $-$ & 20.53/22 & $4.94_{-0.06}^{+0.05}$  \\
X4 & 24.197750 &1 5.795869	& bbody & $-$ & $0.17_{-0.07}^{+0.06}$ & 20.52/16 & $1.92_{-0.07}^{+0.05}$  \\
X5 & 24.173708 & 15.764544  & po  & $1.18_{-0.17}^{+0.21}$ & $-$ & 15.78/14 & $3.98_{-0.07}^{+0.06}$ \\
X8 & 24.132416 & 15.813488 & diskbb &  $-$ & $0.84_{-0.05}^{+0.04}$ & 2.02/3 & $0.92_{-0.14}^{+0.11}$  \\
\hline
\label{T:x-ray_par}
\end{tabular}
\end{table*} 

\begin{table*}
\centering
\caption{The log of {\it HST} observations of NGC 628}
\begin{tabular}{cccccr}
\hline\hline
Instrument & Field  & ObsID & Date & Filter & Exp. \\
 & &  & & &(s)  \\
\hline
ACS/WFC &  Center & J96R23011 & 2005-02-07 & F435W (B) &   1358.0 \\
&  & J96R23021 & 2005-02-07 & F555W (V)   &  858.0 \\
&  & J96R23031 & 2005-02-07 & F658N (H$_{\alpha}$)  &  1422.0 \\
&  & J96R23041 & 2005-02-07 & F814W (I)   &  922.0 \\
\hline
ACS/WFC & East & J96R22R9Q & 2005-06-16 & F435W  &  400.0 \\
&  & J96R22RGQ & 2005-06-16 & F555W &  360.0 \\
&  & J96R22REQ & 2005-06-16 & F814W &  360.0 \\
\hline
ACS/WFC & West  & J96R21011 & 2005-02-16 & F435W  &   1200.0 \\
&  & J96R21021 & 2005-02-16 & F555W   &  1000.0 \\
&  & J96R21031 & 2005-02-16 & F658N   &  1400.0 \\
&  & J96R21041 & 2005-02-16 & F814W   &  900.0 \\
\hline
\end{tabular}
\label{T:obs}
\end{table*}

\begin{table*}
\centering
\caption{Aperture Correction values for the filters of {\it HST}/ACS Wide Field Camera (WFC)}
\begin{tabular}{cccccccc}
\hline\hline
Region & Method & CI & F435W  & F555W &F658N  &F814W  \\
& &&  & &  \\
\hline
& 1& & -0.24 & -0.28 &-& -0.32\\
\hline
& &2.3 < CI < 2.5 &-0.31	& -0.27	&-0.35&	-0.29	\\
Center  & 2 & 2.5 < CI < 2.8 & -0.37 & -0.48 &-0.39&	-0.60	\\
& & CI > 2.8 & -0.59 & -0.52 & -0.47&-0.59		\\
\hline
&1&  & -0.32 & -0.29 &-& -0.34 \\
\hline
& &2.3 < CI < 2.5 & -0.34 & -0.44	&-& -0.18		\\
East & 2 & 2.5 < CI < 2.8 & -0.52	& -0.47 &-& -0.49		\\
& &  CI > 2.8 & -0.40 & -0.58	&-& -0.45	\\
\hline
&1& & -0.35 & -0.39 &-& -0.43\\
\hline
& &2.3 < CI < 2.5 & -0.31 & -0.33 &-0.33& -0.38		\\
West &2 & 2.5 < CI < 2.8 & -0.39 & -0.47	&-0.39& -0.51	\\
& &  CI > 2.8 & -0.43 & -0.48 &-0.44& -0.58		\\
\hline

\end{tabular}
\\Note: The image of the east field taken with the F658N is not available.\\
\label{T:apcor}
\end{table*}

\begin{table*}
\centering
\caption{Coordinates of the X-ray/optical reference sources.}
\begin{tabular}{ccccccccccc}
\hline\hline
$^{*}$Source &	Chandra  R.A	&	Chandra Dec.	&	GAIA R.A	&	GAIA Dec.	& Offsets & \multicolumn{2}{c}{Astrometric Errors}\\
& ($\degr$) & ($\degr$) & ($\degr$) & ($\degr$) & ($\arcsec$) & R.A($\arcsec$) & Decl.($\arcsec$)\\
\hline
2CXO J013641.7 +154701	&	24.17396	&	15.78366	&	24.17394	&	15.78364 & 0.10	\\
2CXO J013637.7+154740	&	24.15713	&	15.79462	&	24.15712	&	15.79462 & 0.01 	\\
& &&&&& 0.4 $\pm$ 0.02 & 0.4 $\pm$ 0.03\\
\hline\hline
Fields &	GAIA R.A	&	GAIA Dec.	&	HST R.A	&	HST Dec.		\\

& ($\degr$) & ($\degr$) & ($\degr$) & ($\degr$) \\
\hline
CENTER	&	24.18498	&	15.76845	&	24.18499	&	15.76845	& 0.03\\
	&	24.15712	&	15.79462	&	24.15713	&	15.79462	& 0.01\\
	&	24.17066	&	15.77894	&	24.17067	&	15.77894	& 0.01\\
	& &&&&& -0.01 $\pm$ 0.011 & -0.01 $\pm$ 0.001\\
									\\
EAST	&	24.22205	&	15.77511	&	24.22204	&	15.77510	& 0.03\\
	&	24.20518	&	15.78055	&	24.20518	&	15.78055	& 0.01\\
	&	24.18325	&	15.77877	&	24.18325	&	15.77877	& 0.03\\
									\\
& &&&&&0.02 $\pm$ 0.01 & 0.02 $\pm$ 0.01\\
WEST	&	24.14788	&	15.80482	&	24.14788	&	15.80482	& 0.03\\
	&	24.14059	&	15.80597	&	24.14058	&	15.80597	& 0.03\\
	&	24.12041	&	15.83753	&	24.12040	&	15.83755	& 0.15\\
	& &&&&& -0.02 $\pm$ 0.07 & -0.04 $\pm$ 0.04\\
\hline
\label{T:astro}
\end{tabular}
\\$^{*}${\it Chandra} Source Catalog Release 2.0 (CSC 2.0).
\end{table*}

\begin{table*}
\centering
\caption{An example list of properties of star clusters in NGC 628.}
\begin{tabular}{ccccccccc}
\hline\hline
ID & R.A & Dec. & B-V  & V-I & r$_{\mathrm{eff}}$ & Age& Mass \\
& &&  &&(pc)&log ($\tau$ /yr) & log (M/M$_{\odot}$)  \\
\hline
1	&	24.202984	&	15.733198	&	0.22	&	0.06	&	4.17	&	6.72	&	3.14	\\
2	&	24.155932	&	15.741452	&	-0.08	&	0.04	&	2.47	&	6.70	&	3.12	\\
3	&	24.154451	&	15.741519	&	0.09	&	1.08	&	2.26	&	6.94	&	3.72	\\
4	&	24.15486	&	15.741673	&	0.13	&	1.03	&	1.88	&	6.95	&	3.00	\\
5	&	24.156314	&	15.741799	&	-0.09	&	0.23	&	9.60	&	6.74	&	2.72	\\
6	&	24.153729	&	15.741869	&	0.20	&	0.42	&	3.27	&	6.85	&	3.07	\\
7	&	24.15578	&	15.742933	&	0.90	&	1.49	&	9.22	&	9.11	&	4.21	\\
8	&	24.156447	&	15.743201	&	0.82	&	1.17	&	8.03	&	9.02	&	4.20	\\
9	&	24.160758	&	15.744161	&	0.32	&	0.74	&	7.23	&	8.31	&	4.46	\\
10	&	24.158929	&	15.744414	&	0.27	&	0.79	&	6.89	&	7.18	&	3.28	\\
... & & & & & & & \\

\hline
\end{tabular}
\label{T:agemass}
\\Notes: In the table (1) the star clusters are ordered according to the Declination (2,3) Right Ascension and Declination in units of degree (4)(B-V) represents (F435W-F555W)$_{0}$ color (5) (V-I) represents (F555W-F814W)$_{0}$ color (6) effective radius (radius containing half of the total cluster light) (7) age of the cluster calculated from BVI colors (8) mass of the cluster.
\end{table*}

\begin{table*}
\centering
\caption{X-ray and optical properties of XRBs in NGC 628}
\begin{tabular}{c c c c c c c c c c c}
\hline \hline
ID	&	R.A. &	Dec. &	V$_{f}$	& v/t	& optical	& ID of cluster & Cluster Age & Cluster Mass & Dist. & Classification	\\
	&	 & 	&	&  &	counterpart &   & log ($\tau$ /yr) & log (M/M$_{\odot}$)& (pc)	&	\\
(1) & (2) & (3) & (4) & (5) & (6) & (7) & (8) & (9) & (10) &(11) \\
	
\hline
X1	&	24.21284	&	15.76308	&	50	&	v	&	y	&		&		&		& & HMXB		\\
X2	&	24.18388	&	15.80520	&	256	&	t	&	y	&		&		&		&	& HMXB	\\
X5	&	24.17371	&	15.76454	&	460	&	t	&	y	&	183	&	8.09	&	3.73	&	194 & IMXB	\\
X8	&	24.13239	&	15.81348	&	266	&	t	&	n	&		&		&		&	& LMXB	\\
X14	&	24.18287	&	15.79925	&	20	&	t	&	n	&		&		&		&	& LMXB	\\
X16	&	24.21795	&	15.79370	&	18	&	t	&	y	&	565 (e)	&	8.60	&	4.60	&	169	& HMXB\\
X20	&	24.16418	&	15.79564	&	24	&	t	&	y	&	602	&	6.74	&	3.53	&	56	& HMXB\\
X21	&	24.14906	&	15.76563	&	5	&	v 	&	n	&		&		&		&	& LMXB	\\
X22	&	24.20465	&	15.75766	&	4	&	uc	&	y	&		&		&		&	& HMXB	\\
X23	&	24.16268	&	15.79876	&	6	&	t$_{c}$	&	n	&	649	&	8.90	&	3.97	&	49 & LMXB	\\
X24	&	24.18727	&	15.76285	&	3	&	uc	&	y	&	142	&	6.76	&	3.35	&	138	& HMXB \\
X28	&	24.17308	&	15.78055	&	132	&	t	&	n	&	400	&	10.03	&	6.10	&	123	 & LMXB \\
X30	&	24.18032	&	15.78606	&	9	&	t	&	n	&		&		&		&	& LMXB	\\
X32	&	24.16937	&	15.78701	&	6	&	t$_{c}$	&	n	&	484	&	9.34	&	4.84	&	71	& LMXB \\
X33	&	24.15700	&	15.78822	&	23	&	t	&	n	&	501	&	8.40	&	3.82	&	109	& LMXB \\
X34	&	24.18473	&	15.77495	&	7	&	t$_{c}$	&	y	&		&		&		&	& IMXB	\\
X35	&	24.20312	&	15.77867	&	82	&	t	&	n	&		&		&		&	& LMXB	\\
X37	&	24.13160	&	15.80653	&	64	&	t	&	n	&	757	(w) &	6.90	&	2.78	&	167	& LMXB \\
X38	&	24.12328	&	15.78241	&	74	&	t	&	n	&		&		&		&	& LMXB	\\
X40	&	24.18196	&	15.79524	&	9	&	t$_{c}$	&	n	&	595	&	9.24	&	3.04	&	89	& LMXB\\
X41	&	24.19433	&	15.76273	&	32	&	t	&	n	&		&		&		&	& LMXB	\\
X42	&	24.16700	&	15.77371	&	50	&	t	&	n	&		&		&		&	& LMXB	\\
X45	&	24.16954	&	15.77980	&	67	&	t	&	n	&		&		&		&	& LMXB	\\
X46	&	24.18856	&	15.79659	&	29	&	t	&	n	&	615	&	6.75	&	5.08	&	6 & LMXB	\\
X49	&	24.20032	&	15.74579	&	6	&	t$_{c}$	&	n	&		&		&		&	& LMXB	\\
X50	&	24.20346	&	15.78153	&	34	&	t	&	n	&		&		&		&	& LMXB	\\
X52	&	24.22701	&	15.76107	&	6	&	t$_{c}$	&	n	&		&		&		&	& LMXB	\\
X55	&	24.12527	&	15.81537	&	58	&	t	&	n	&	807 (w)	&	6.74	&	3.08	&	61 & LMXB	\\
X56	&	24.14774	&	15.83150	&	70	&	t	&	y	&		&		&		&	& HMXB	\\
X57	&	24.14701	&	15.78253	&	7	&	t$_{c}$	&	n	&		&		&		&	& LMXB	\\
X58	&	24.16352	&	15.76690	&	63	&	t	&	y	&		&		&		&	& HMXB	\\
X63	&	24.23284	&	15.77626	&	58	&	t	&	n	&		&		&		&	& LMXB	\\
X64	&	24.22136	&	15.75302	&	17	&	t	&	n	&		&		&		&	& LMXB	\\
X65	&	24.15804	&	15.79704	&	54	&	t	&	y	&		&		&		&	& HMXB	\\
X66	&	24.14719	&	15.78662	&	55	&	t	&	y	&	476	&	6.85	&	3.28	&	198	& HMXB \\
X68	&	24.16532	&	15.79690	&	24	&	t	&	n	&	631	&	8.91	&	3.74	&	163	& LMXB\\
X69	&	24.17684	&	15.78372	&	53	&	t	&	y	&		&		&		&	& HMXB	\\
X70	&	24.15051	&	15.75310	&	16	&	t	&	n	&	53	&	8.51	&	4.01	&	140	& LMXB\\
X72	&	24.21031	&	15.78476	&	13	&	t	&	y	&		&		&		&	& HMXB	\\
X73	&	24.12003	&	15.81654	&	27	&	t	&	n	&	812	(w) &	9.11	&	4.98	&	29 & LMXB	\\
X74	&	24.19622	&	15.76194	&	29	&	t	&	y	&	132	&	6.80	&	3.21	&	48 & HMXB	\\
X75	&	24.17206	&	15.75149	&	11	&	t	&	n	&	35	&	8.76	&	4.12	&	187	& LMXB\\

\hline
\end{tabular}
\label{T:x-optic}
\\Notes:(1) ID of XRB candidates (2,3) {\it Chandra} positions in units of degree. (4) The source variability V$_{f}$=F$_{max}$/F$_{min}$ is given by the ratio of the maximum and minimum flux in 0.3-8 keV. (5) v: Variable sources, t: transient sources, t$_{c}$: transient candidates, uc: unclear (6) yes or no (y/n) whether the X-ray sources have optical counterparts within the astrometric error radius or not. (7) ID of associated star clusters with XRB. (w), (e) and the rest means west, east and center fields, respectively. (8) age of the cluster calculated from BVI colors (9) mass of the cluster (10) Distance between cluster and XRB in units of parsec (11) Classification of XRBs high/low mass based on their optical counterparts.  \\
\end{table*}

\begin{table*}
\centering
\caption{Probability of chance alignment with XRBs for nearby clusters.}
\begin{tabular}{lccc}
\hline\hline
ID & $\sigma$ & $\Delta$ & 1 $-$ P$_{un}$($\sigma$, $\Delta$)  \\
 & (arcsec$^{-2}$) x 10$^{-3}$ & (arcsec) & (\%)  \\
\hline
X5 & 13.63 & 4.27 & 45.8 \\
X16(e) & 1.53 & 3.69 & 93.7 \\
X20 & 2.08 & 1.29 & 98.9 \\
X23 & 2.29 & 1.11 & 99.1 \\
X24 & 3.97 & 2.99 & 89.4 \\
X28 & 0.63 & 2.73 & 98.5 \\
X32 & 6.66 & 1.59 & 94.8 \\
X33 & 10.86 & 2.42 & 81.9 \\
X37(w) & 5.05 & 3.78 & 79.7 \\
X40 & 10.86 & 1.91 & 88.2 \\
X46 & 0.03 & 0.14 & 99.9 \\
X55(w) & 1.39 & 1.41 & 99.1 \\
X66 & 4.84 & 4.17 & 76.7 \\
X68 & 10.86 & 3.49 & 66.0 \\
X70 & 12.31 & 3.02 & 70.2 \\
X73(w) & 0.47 & 0.63 & 99.9 \\
X74 & 4.84 & 1.03 & 98.4 \\
X75 & 12.31 & 4.02 & 53.5 \\
\hline
\end{tabular}
\label{T:prob}
\\Notes: (w), (e) and the rest means west, east and center fields, respectively. \\
\end{table*}

\end{document}